\documentclass[12pt]{report}
\usepackage[utf8]{inputenc}
\usepackage[T1]{fontenc}
\usepackage{svg}
\usepackage{graphicx}
\graphicspath{{images/}}
\usepackage[a4paper, width=150mm, top=25mm, bottom=25mm, bindingoffset=6mm]{geometry}
\usepackage{parskip}
\usepackage{caption}
\usepackage{subcaption}
\usepackage{authblk}
\usepackage{float}
\usepackage{enumitem}
\usepackage[mathletters]{ucs}
\usepackage{comment}
\usepackage[section]{placeins}
\usepackage{blindtext}
\usepackage{amsmath}
\usepackage{amssymb}
\usepackage{changepage}
\usepackage{array}
\usepackage{longtable}
\usepackage{xcolor}
\usepackage{relsize}
\usepackage[hidelinks]{hyperref}
\usepackage{apacite}

\usepackage{algorithm2e}
\RestyleAlgo{ruled}

\SetCommentSty{mycommfont}

\title{\textbf{Model-Free Reinforcement Learning for Asset Allocation} \\  [0.5em]\smaller{} \vspace{20pt} Practicum Final Report}

\author{Adebayo Oshingbesan}
\author{Eniola Ajiboye}
\author{Peruth Kamashazi}
\author{Timothy Mbaka}
\affil{Carnegie Mellon University Africa}

\author{Mahmoud Mahfouz}
\author{Sood Srijan}
\affil{JP Morgan}

\author{David Vernon}
\affil{Carnegie Mellon University Africa}

\date{\today}

\begin{document}
\begin{titlepage}

\newcommand{\HRule}{\rule{\linewidth}{0.5mm}} 


\begin{figure}[htbp]
  \centering
  \includegraphics[width=0.4\textwidth]{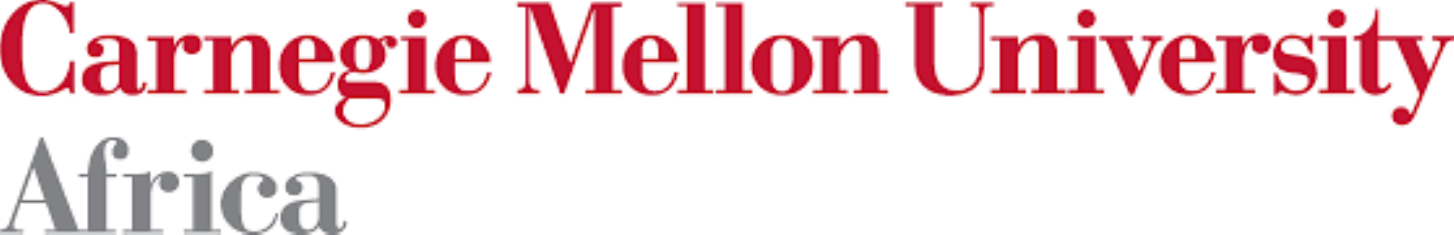} 
  \hfill
  \includegraphics[width=0.3\textwidth]{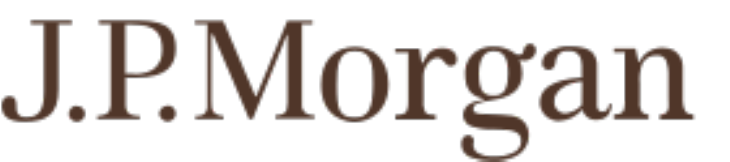}
\end{figure} 

\vspace{4.25cm}

\center 



\makeatletter
\HRule \\[0.4cm]
{ \huge \bfseries Model-Free Reinforcement Learning for Asset Allocation}\\[0.4cm] 
\HRule \\[1.5cm]

\textsc{\LARGE Practicum Final Report}\\[1.5cm] 
 
\vspace{4.25cm}

\begin{minipage}{0.4\textwidth}
\begin{flushleft} \large
\emph{Authors:}\\
Adebayo Oshingbesan
\vspace{0.025cm}\\    
Eniola Ajiboye
\vspace{0.025cm}\\    
Peruth Kamashazi
\vspace{0.025cm}\\    
Timothy Mbaka
\vspace{0.025cm}\\ 
\end{flushleft}
\end{minipage}
~
\begin{minipage}{0.4\textwidth}
\begin{flushright} \large
\emph{Industry Advisor \& Client:} \\
Mahmoud Mahfouz
\vspace{0.025cm}\\    
Srijan Sood \\[0.8em] 
\emph{Faculty Supervisor:} \\
David Vernon 
\vspace{0.025cm}\\ 
\end{flushright}
\end{minipage}\\[2cm]
\makeatother



{\large \today}\\[2cm] 

\vfill 

\end{titlepage}




\chapter*{Acknowledgments}
This work would not have been possible without the advice and support of several people. First and foremost, we would like to express our gratitude to Mahmoud Mahfouz, Vice President of AI and Research at J.P. Morgan and Chase Co., and his colleague Sood Srijan for providing us with resources and valuable guidance during this project. We would also like to thank our advisor, Prof. David Vernon, for his guidance and constant support.

\renewcommand{\contentsname}{Table of Contents}
\tableofcontents

\listoffigures

\listoftables

\chapter{Introduction}
\section{Background}
Asset allocation (or portfolio management) is the task of determining how to optimally allocate funds of a finite budget into a range of financial instruments/assets such as stocks \cite{filos_reinforcement_2019}. Coming up with a profitable trading strategy involves making critical decisions on allocating capital into different stock options. Usually, this allocation should maximize the expected return while minimizing the investment risks involved \cite{gao_framework_2021}. There are several existing portfolio management strategies, and the state-of-the-art portfolio management frameworks are broadly classified into baseline models, follow the winner, follow the loser, pattern-matching, and meta-learning\cite{li_online_2014}.

While many of these state-of-the-art models achieve good results, there are some limitations. First, they are overly reliant on using predictive models \cite{adammer2020forecasting}. These predictive models are not usually too successful at predicting the financial markets since these markets are highly stochastic and thus are very difficult to predict accurately\cite{belousov_reinforcement_2021}. Similarly, many of these models make simplistic and usually unrealistic assumptions around the financial signals' second-order and higher-order statistical moments \cite{gao_framework_2021}. Finally, these models are usually limited to discrete action spaces to make the resulting models tractable to solve \cite{filos_reinforcement_2019}.

Reinforcement learning (RL) has become increasingly popular in financial portfolio management \cite{huang_deep_2020}. Reinforcement learning is the sub-field of machine learning concerned with how intelligent machines ought to make decisions in an environment to maximize the cumulative reward over time \cite{noauthor_reinforcement_2021}. The addition of deep neural networks have been one of the breakthrough concepts in reinforcement learning in recent years, achieving superhuman performance in several tasks such as playing chess \cite{silver_general_2018}.

\section{Problem Definition}
While there are a lot of tools currently available for portfolio management, they are limited in their abilities because of inherent assumptions about the financial markets. Techniques that reduce the number of simplifying assumptions made could yield better results. Thus, there may be a significant gap between what is possible and what is currently available in the field.

\section{Aim and Objectives}
This study investigates the effectiveness of model-free deep reinforcement learning agents with portfolio management. The specific objectives are:
\begin{itemize}
    \item Training RL agents on real-world prices of a finite set of stocks to optimally allocate a finite cash budget into a range of securities in a portfolio.
    \item Comparing the performance of RL agents to baseline agents.
    \item Comparing the performance of value-based RL agents to policy-based RL agents.
    \item Comparing the performance of off-policy RL agents to on-policy RL agents.
\end{itemize}

\section{Research Questions}
At the end of this report, we should be able to answer the following questions:
\begin{enumerate}
    \item How well can RL agents perform the task of portfolio management?
    \item Are RL agents markedly better than the classical state-of-the-art portfolio management techniques?
    \item Are there certain classes of RL agents that are consistently better at portfolio management?
\end{enumerate}

\section{Significance of Study}
Since Harry Markowitz proposed the idea of portfolio formation in 1952 to improve expected rewards and reduce the investment risk, portfolio trading has been the dominant strategy for many traders. Any technique that helps improve this strategy will have many positive ripple effects \cite{levisauskait_investment_nodate-1}. Reinforcement learning has shown that it is capable of improving performance in certain tasks substantially. An excellent example of this is the game of chess where AlphaZero and LeelaZero, two RL agents, beat state-of-the-art chess engines. These RL agents won by using the strategy of long-term positional advantage over materialism leading many chess players to modify their playing styles \cite{silver_general_2018}. This study aims to understand if RL can uncover strategies for portfolio management that yield better results than the current state of the art methods.

\section{Limitations of Study}
While this study will not be making any strong assumptions regarding the financial signals obtained from the market, it will make some reasonable assumptions around the impact of the RL agent on the market.  However, these assumptions are generally considered reasonable and should not impact the portability of this study into the real world.

\section{Structure of Report}
The remainder of this report is structured as follows:
\begin{itemize}
    \item Chapter 2: Portfolio Management - This chapter provides an overview of the field of portfolio management.
    \item Chapter 3: Survey of Machine Learning in Finance - This chapter provides an overview of the different applications of machine learning in finance, especially in reinforcement learning for portfolio management.
    \item Chapter 4: Financial Environment - This chapter describes how the trading environment is modeled as a reinforcement learning environment.
    \item Chapter 5: Survey of Reinforcement Learning Techniques - This chapter provides an overview of the field of reinforcement learning.
    \item Chapter 6: Trading Agents - This chapter describes the baseline and RL agents considered in this work.
    \item Chapter 7: Experiments - This chapter describes all the experiments that were carried  in this study.
    \item Chapter 8: Results and Discussion - This chapter documents and discusses the results of this study. 
    \item Chapter 9: Conclusion - This chapter summarizes the results, answers the initial research questions, ensures that the study aim \& objectives were achieved, and recommends future directions for the study.
\end{itemize}

\chapter{Portfolio Management}
Portfolio management involves selecting and managing a collection of assets in order to meet long-term financial goals while adhering to a risk tolerance level. Diversification and portfolio optimization are critical components of efficient portfolio management. Diversification is a risk management approach that involves combining a wide range of investments in a portfolio. A diversified portfolio comprises various asset types and investment vehicles to reduce exposure to any particular asset or risk \cite{Hayes2021-ig}.

\section{Portfolio Optimization}
A portfolio is a collection of several financial assets. Portfolio optimization is the process of determining the optimum portfolio from a set of all possible portfolios with a specific goal in mind. An optimum portfolio mixes candidate assets so that the chance of the portfolio generating a positive return is maximized for a given level of risk  \cite{Yiu2020-mx}. Portfolio optimization systematically solves the asset allocation problem by constructing and optimizing an objective function expressing the investor’s preferences concerning the portfolio vector \cite{filos_reinforcement_2019}.

One method for accomplishing this is to assign equal weights to each asset. When the returns across the assets are random and there is no historical data, assigning equal weights is considered the most basic type of asset allocation and can be effective. This strategy, however, is inefficient since it does not take into account past data. Thus, more efficient methods have been developed. One of these methods is the Markowitz Model.

\section{Markowitz Model}
The Markowitz model \cite{10.2307/2975974} is regarded as one of the first efforts to codify and propose an optimization strategy for portfolio management mathematically. The Markowitz model formulates portfolio allocation as discovering a portfolio vector $w$ among a universe of $M$ assets. The Markowitz model gives the optimal portfolio vector $w *$ which minimizes volatility for a given return level, such that \cite{filos_reinforcement_2019}:
\begin{equation}
        \sum_{i=1}^{M} w_{*,i} = 1,  w_x \in\mathbb{R}
    \label{markowitz equation}
\end{equation}

The Markowitz model assumes that the investor is risk-averse and thus determines the optimal portfolio by selecting a portfolio that gives a maximum return for a given risk or minimum risk for given returns. Therefore, the optimal portfolio is selected as follows:
\begin{itemize}
    \item From a set of portfolios with the same return, the investor will prefer the portfolio with the lower risk.
    \item From a set of portfolios with the same risk, the investor will choose the portfolio with the highest return.
\end{itemize}

The Markowitz model also presupposes that the analysis is based on a one-period investment model and is thus only applicable in a one-period situation. This means that the choice to distribute assets is made only at the start of the term. As a result, the repercussions of this decision can only be seen after the term, and no additional action may be taken during that time. This makes the model a static model. However, since the Markowitz model is simple to grasp, it is frequently utilized as the cornerstone for current portfolio optimization strategies due to its simplicity and efficiency. However, because it is based on a single-period investment model, it is not ideal for continuous-time situations.

\section{Modern Portfolio Theory}
The Markowitz model serves as the foundation for the modern portfolio theory (MPT). MPT is a mathematical framework for constructing a portfolio of assets to maximize the expected return for a given amount of risk. Diversification is an essential component of the MPT. Diversification refers to the concept that having a variety of financial assets is less hazardous than owning only one type \cite{Silver2021-ab}.

A core principle of the modern portfolio theory is that an asset's risk should be measured not by itself but by how it contributes to the total risk and return of the portfolio. Modern portfolio theory uses the standard deviation of all returns to assess the risk of a specific portfolio \cite{E2021-zr}. Modern portfolio theory may be used to diversify a portfolio in order to get a higher overall return with less risk. The efficient frontier is a key concept in MPT. It is the line indicating the investment combination that will deliver the best level of return for the lowest degree of risk \cite{Silver2021-ab}.

\section{Post-modern Portfolio Theory}
The post-modern portfolio theory (PMPT) extends the modern portfolio theory (MPT). PMPT is an optimization approach that uses the downside risk of returns rather than the expected variance of investment returns employed by MPT. The difference in risk between the PMPT and the MPT, as measured by the standard deviation of returns, is the most important aspect of portfolio creation. The MPT is based on symmetrical risk, whereas the PMPT is based on asymmetrical risk. The downside risk is quantified by target semi-deviation, also known as downside deviation, and it reflects what investors dread the most: negative returns \cite{Chen2021-pq}.

\chapter{Survey of Machine Learning in Finance}
\section{Introduction}
\begin{figure}[ht]
    \centering
    \includegraphics[scale=0.6]{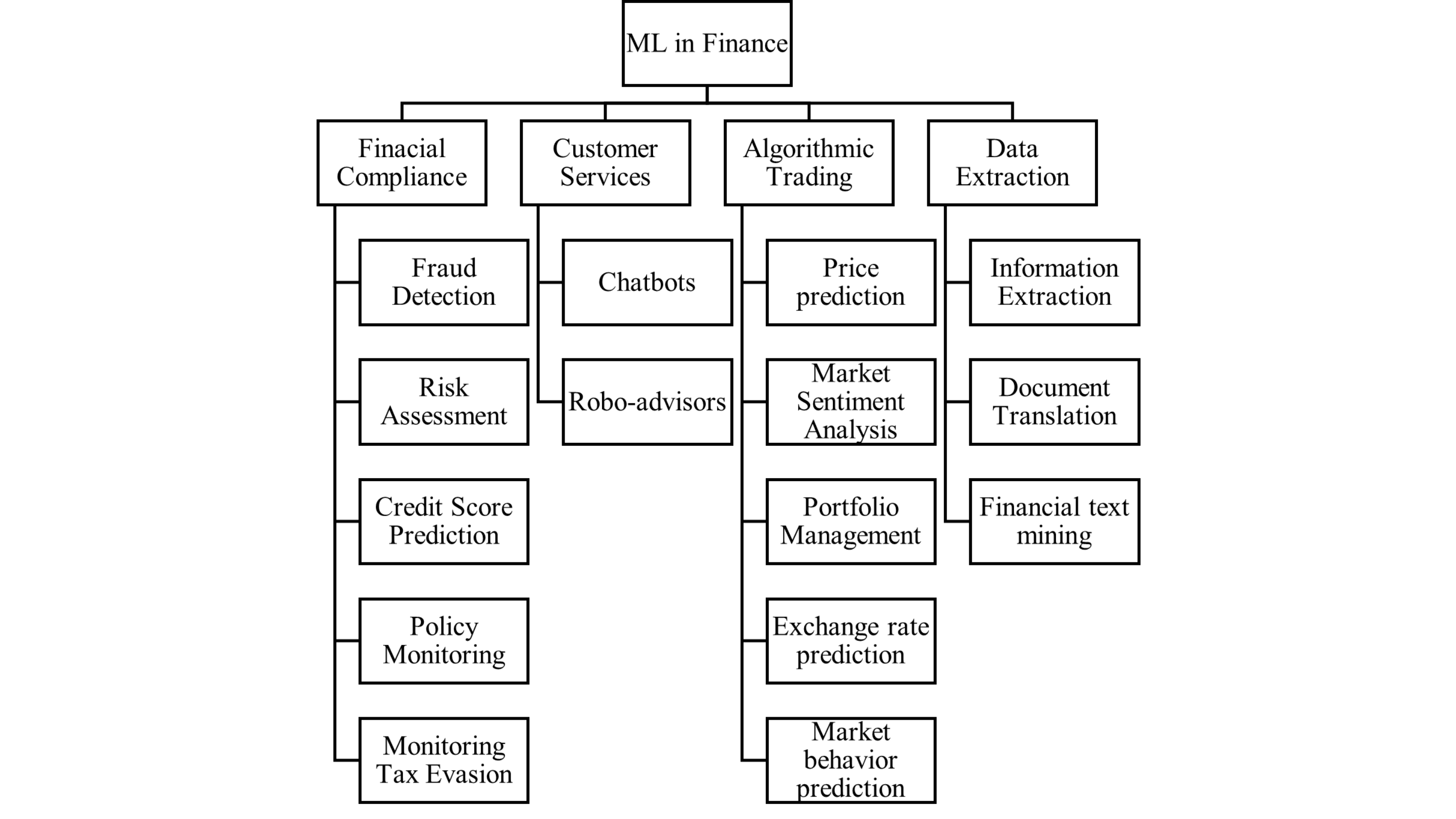}
    \caption{Taxonomy of Common Applications of Machine Learning in Finance}
    \label{fig: ML Taxonomy in Finance}
\end{figure}

Machine learning has become increasingly important in the finance industry across several tasks. Figure \ref{fig: ML Taxonomy in Finance} shows a taxonomy of some of the typical applications of machine learning in finance.

For financial tasks, different classes of machine learning algorithms have been frequently used. These classes include generalized linear models, tree-based models, kernel-based models, and neural networks. Similarly, the data has widely varied from structured data such as tables to unstructured data such as text. Table \ref{tab: previous research} itemizes various finance research articles where some of these models were used.

\begin{longtable}{| >{\raggedright}p{0.4\textwidth} | 
>{\raggedright}p{0.2\textwidth} | >{\raggedright}p{0.2\textwidth} | >{\raggedright\arraybackslash}p{0.2\textwidth} | }

    \caption{Table of Previous ML Research in Finance.} \\
    \hline
    \hline
    \textbf{Reference} & \textbf{Task} & \textbf{Data} & \textbf{ML Algorithm(s)} \\
    \hline
    \hline
    \cite{adammer2020forecasting} & Information extraction and Risk assessment (equity risk premium prediction) & News articles & Correlated Topic Model (CTM), LASSO, Support Vector Regression, Ridge Regression, Elastic Net, Decision Trees, Random Forests, Boosted Trees \\
    \hline
    \cite{albanesi2019predicting} & Credit default behavior & Credit bureau data & Boosted Trees and Deep Neural Networks \\
    \hline
    \cite{amel2020machine} & Abnormal Stock Returns & Financial Statement Variables & LASSO, Random Forests, Neural Networks \\
    \hline
    
    \cite{ang2020using} & Start-ups Valuations and Probability of Success & Start-up funding data and descriptions from Crunchbase over ten years, & LDA and Gradient Boosting Regressor \\
    \hline
    
    \cite{antweiler2004all} & Stock market sentiment analysis & Yahoo Finance message board & Naïve Bayes and SVM \\
    \hline
    
    \cite{bao2017deep} & Stock prices forecasting & Technical Indicators & LSTM and Stacked Autoencoders \\
    \hline
    
    \cite{bao2020detecting} & Accounting Fraud & Raw financial statement & Boosted Trees \\
    \hline
    
    \cite{bari2020ensembles} & Trading signal generations & Tweets and Financial News & LSTMs and GRUs \\
    \hline
    
    \cite{bjorkegren2020behavior} & Payment of bills – credit risk prediction & Mobile phone metadata & Random Forests \\
    \hline
    
    \cite{chen2020deep} & Stochastic discount factor & Firm characteristics, historical return, and macroeconomic indicators. & Generative Adversarial Networks \\
    \hline
    
    \cite{colombo2020statistical} & The direction of changes in exchange rates & Marker uncertainty indicator data & Support Vector Machines \\
    \hline
    
    \cite{croux2020important} & Loan Default & Loan data, borrowers’ characteristics, macroeconomic indicators & LASSO \\
    \hline
    
    \cite{gomes2017identifying} & Anomaly detection & Chamber of Deputies open data, Companies data from Secretariat of Federal Revenue of Brazil & Deep Autoencoders \\
    \hline
    
    \cite{goumagias2018using} & Tax evasion prediction & Empirical data from Greek firms & Deep Q-Learning \\
    \hline

    \cite{gu2020empirical} & Stock returns forecasting & Firm characteristics, Historical returns, Macroeconomic indicator & Linear Regression, Random Forests, Boosted Trees, Neural Networks \\
    \hline
    
    \cite{gulen2020application} & Estimation of heterogeneous treatment effects of debt covenant violations on firms’ investment levels. & Firms characteristic data & Causal Forests \\
    \hline
    
    \cite{huang2016exploiting} & Price direction prediction & Tweets & Neural Networks \\
    \hline
    
    \cite{iwasaki2018topic} & Stock price prediction & Analyst Report & LSTM, CNN, Bi-LSTM \\
    \hline
    
    \cite{jiang2017cryptocurrency} & Cryptocurrency portfolio management & Cryptocurrency price data & RNN, LSTM, CNN \\
    \hline
    
    \cite{kvamme2018predicting} & Mortgage default prediction & Mortgage data from Norwegian financial service group, DNB & Convolutional Neural Networks and Random Forest \\
    \hline
    
    \cite{lahmiri2019can} & Corporate Bankruptcy & Firms’ financial statements, market data, and general risk indicators & Neural networks \\
    \hline
    
    \cite{li2018stock} & Stock price prediction & Stocks data & RNN, LSTM, GRU \\
    \hline
    
    \cite{liang2018adversarial} & Portfolio Allocation & Stocks data & Deep Reinforcement Learning \\
    \hline

    \cite{luo2017deep} & Corporate credit rating & CDS data & Deep Belief Network and Restricted Boltzmann Machines \\
    \hline
    
    \cite{ozbayoglu2020deep} & Financial distress prediction & Financial News & SVM, Deep Belief Network, LSTM \\
    \hline
    
    \protect\cite{damrongsakmethee2017data} & Credit score classification & Credit scores data & MLP and CNN \\
    \hline
    
    \cite{paula2016deep} & Financial fraud and money laundering & Databases of foreign trade of the Secretariat of Federal Revenue of Brazil & Deep Autoencoders \\
    \hline
    
    \cite{reichenbacher2020expected} & future bond liquidity & Bond transactions and characteristics data & Elastic Nets and Random Forests \\
    \hline
    
    \cite{antunes2017probabilistic} & Bankruptcy Prediction & Financial Statements & Deep Belief Network \\
    \hline
    
    \cite{rossi2020benefits} & Investors portfolio allocation and performance and effects of Robo-advising & Investor characteristics & Regression Trees \\
    \hline
    
    \cite{ozbayoglu2020deep} & Credit card default & Account data and macroeconomic indicators & Decision Trees, Random Forest, Boosted Trees \\
    \hline
    
    \cite{taghian2020learning} & Trading Signal Generation & Market data & Neural Network, Genetic Programming, and Reinforcement Learning \\
    \hline
    
    \cite{spilak2018deep} & Dynamic portfolio allocation & Cryptocurrency data & LSTM, RNN, MLP \\
    \hline
    
    \cite{ozbayoglu2020deep} & Stock classification & Stocks data & Deep RBM Encoder-Classifier Network \\
    \hline
    
    \cite{tian2015variable} & Corporate Bankruptcy & Firms’ financial statements and market data & LASSO \\
    \hline
    
    \cite{vamossy2021investor} & Investor’s emotions & StockTwits post & Deep Neural Networks \\
    \hline
    
    \cite{deng2016hierarchical} & Stock price prediction and trading signal generation. & Stock price data & Fuzzy Deep Direct Reinforcement Learning \\
    \hline \hline

\label{tab: previous research}
\end{longtable}

\section{RL Applications in Portfolio Management}
As seen in table \ref{tab: previous research}, reinforcement learning has been applied in stock price prediction, portfolio management/allocation, tax evasion prediction, and trading signal generation, among others. In the following paragraphs, we will be providing an overview of several significant works that have been done in the domain of reinforcement learning for portfolio management in the past few years.

A study proposed an extendable reinforcement learning framework for handling a generic portfolio management challenge \cite{jiang_deep_2017}. At the lowest level, the framework is based on the Ensemble of Identical Independent Evaluators (EIIE) meta topology, which allows for many different forms of weight-sharing neural networks. This framework was tested on the bitcoin market, and it outperformed other trading algorithms by a wide margin.

Another study \cite{filos_reinforcement_2019} introduced a global model-free reinforcement learning family of agents. Because it generalizes across assets and markets independent of the training environment, this methodology proved economical memory-wise and computation-wise. Furthermore, the author used pre-training, data augmentation, and simulation to ensure more robust training.

A group of researchers \cite{huotari_deep_2020} studied the portfolio performance of a trading agent based on a convolutional neural network model. The agent's activity correlated with an investor's high risk-taking behavior. Furthermore, the agent beat benchmarks, although its performance did not differ statistically substantially from the S{\&}P 500 index.

In deep reinforcement learning for portfolio management \cite{hieu_deep_2020}, the authors examined three cutting-edge continuous policy gradient algorithms - deep deterministic policy gradient (DDPG), generalized deterministic policy gradient (GDPG), and proximal policy optimization (PPO). The authors concluded that the first two performed significantly better than the third.

A research work \cite{gao_framework_2021} suggested a hierarchical reinforcement learning framework that can manage an arbitrary number of assets while accounting for transaction fees. On real-world market data, the framework performs admirably. However, since just four equities were evaluated, there is some doubt on the framework's capacity to deal with enormous amounts of data.

\chapter{Financial Environment}
\section{Assumptions}
The real-world financial market is a highly complex system. In this work, we model the financial environment as a discrete-time, stochastic dynamic system with the following simplifying assumptions:
\begin{itemize}
    \item There is no dependence on explicit stock price prediction (model-free).
    \item The actions of the RL agent should be continuous.
    \item There will be zero slippage.
    \item The RL agents have zero market impact.
    \item Short-selling is prohibited.
    \item The environment is a partially observable system.
\end{itemize}

These simplifying assumptions are consistent with similar works in literature (\cite{filos_reinforcement_2019}, \cite{gao_framework_2021}, \cite{betancourt_deep_2021}, \cite{jiang2017cryptocurrency}, \cite{hu_deep_2019}).

\section{Description}
The environment takes in the following inputs:
\begin{itemize}
        \item Data: This is either a dataframe or a list of stock tickers. If a dataframe is provided, the index of the dataframe must be of type datetime (or can be cast to datetime). Each column should contain the prices of the stock name provided in the header over the time period.
               
        \item Episode Length: This refers to how long (in days) the agent is allowed to interact with the environment. 

        \item Returns: The environment has two reward signals (see section 4.4). The returns variable is a boolean flag used to choose between these reward signals. When set to true, the environment uses the log-returns as the reward signal. When set to false, the environment used the differential Sharpe ratio.

        \item Trading Cost Ratio: This is the percentage of the stock price that will be attributed to the cost of either selling or buying a unit stock.

        \item Lookback Period: This is a fixed-sized window used to control how much historical data to return to the agent as the observation at each timestep.

        \item Initial Investment: This refers to the initial amount available to the agent to spend on all the available stocks in the environment.

        \item Retain Cash: This is a boolean flag value used to inform the environment whether the agent can keep a cash element or not.

        \item Random Start Range: The agent is encouraged to start from a random range to avoid over-fitting. This value controls what that range should be.

        \item DSR Constant: This is a smoothing parameter for the differential Sharpe ratio.

        \item Add Softmax: This is a boolean flag that controls whether the environment should perform the softmax operation or not. This is required to support the out-of-the-box RL agents from other libraries.

        \item Start Date: If a list of tickers was provided instead of a dataframe, this start date parameter is used for the yahoo finance data download.

        \item End Date: If a list of tickers was provided instead of a dataframe, this end date parameter is used for the yahoo finance data download.

        \item Seed: This is a seed value for environment reproducibility
\end{itemize}

\section{State Space}
At any time step t, the agent will observe a stack of T vectors such that T is the amount of lookback context that the agent can observe.
Each of the vectors will be an asset information vector denoted as $P_t$ where:

\begin{equation}
            P_t=\left[\ \log{\left(\frac{P_1,_t}{P_{1,t-1}}\right)},\ \log{\left(\frac{P_2,_t}{P_{2,t-1}}\right)},\ \ldots,\ \log{\left(\frac{P_M,_t}{P_{M,t-1}}\right)}\right]^T\ \ \epsilon\ R^M
\end{equation}

\section{Action Space}
The action represents the weights of the stocks in a portfolio at any time t where its i-th component represents the ratio of the i-th asset such that:

    \begin{equation} \label{action_space_1}
    A_t=\left[A_t,_1\ ,\ A_t,_2\ ,\ \ldots,\ A_t,_M\right]^T\ \epsilon\ R^M
    \end{equation}
    \begin{equation} \label{action_space_2}
    	\sum_{i=1}^{M}{A_i,_t=1}
    \end{equation}
    \begin{equation} \label{action_space_3}
    	{0\le A}_i,_t\le1\ \forall\ i,t\   
    \end{equation}
    where M\ is\ the\ total\ number\ of\ assets\ in\ the\ portfolio

If the agent is required to keep a cash element, the weight vector's size is increased by one yielding an extended portfolio vector that satisfies equations \ref{action_space_1} to \ref{action_space_3} with the cash element treated as an additional asset.

\section{Reward functions}
There are two reward functions $R_t$ - log-returns and differential Sharpe ratio. The specific training reward returned for a particular training episode will be determined by the returns input value to the environment. The reward functions are defined as follows:

\begin{itemize}
\item \textbf{Log-returns} \newline
This is defined as the weighted sum of log-returns for the portfolio such that:
\begin{equation}
    R_t={In(1\ +\ P}_{t+1}\bullet A_t)\ \ 
\end{equation}

\item \textbf{Differential Sharpe Ratio} \newline
This is defined as an instantaneous risk-adjusted Sharpe ratio. Equations \ref{sr_1} to \ref{sr_n} provide its mathematical formulation \cite{filos_reinforcement_2019}.

\begin{equation}\label{sr_1}
    R_t= (Y_{t-1} * \delta X_t+0.5X_{t-1} * \delta Y_t)/(Y_t-1 - X{^2}_{t-1})^{1.5}   
\end{equation}

\begin{equation}
    X_t=X_{t-1}+\ \alpha * \delta X_t
\end{equation}

\begin{equation}
    Y_t=Y_{t-1}+\ \alpha * \delta Y_t
\end{equation}

\begin{equation}
    \delta X_t=LR_t- X_{t-1}
\end{equation}

\begin{equation} \label{sr_n}
    \delta Y_t=LR{^2}_{t}- Y_{t-1} ,
\end{equation}

where: \newline
LR is log-returns \newline
$0 <= \alpha\ <=1; \alpha\ is\ a\ smoothing\ factor.$

\end{itemize}

\chapter{Survey of Reinforcement Learning Techniques}
\section{Introduction to Reinforcement Learning}
Reinforcement learning (RL) involves learning by interaction with an environment led by a specified goal. The agent learns without being explicitly programmed, selecting actions based on prior experience \cite{huang_deep_2020}. RL is growing more prominent as computer technologies such as artificial intelligence (AI) have advanced over the years. Machine learning (ML) is a sub-field of AI that focuses on the use of data and algorithms to emulate the way humans learn, eventually improving its performance\cite{filos_reinforcement_2019}. There are mainly three categories of ML, namely supervised learning, unsupervised learning, and reinforcement learning. 

\begin{figure}[!ht]
    \centering
    \includegraphics[scale=0.5]{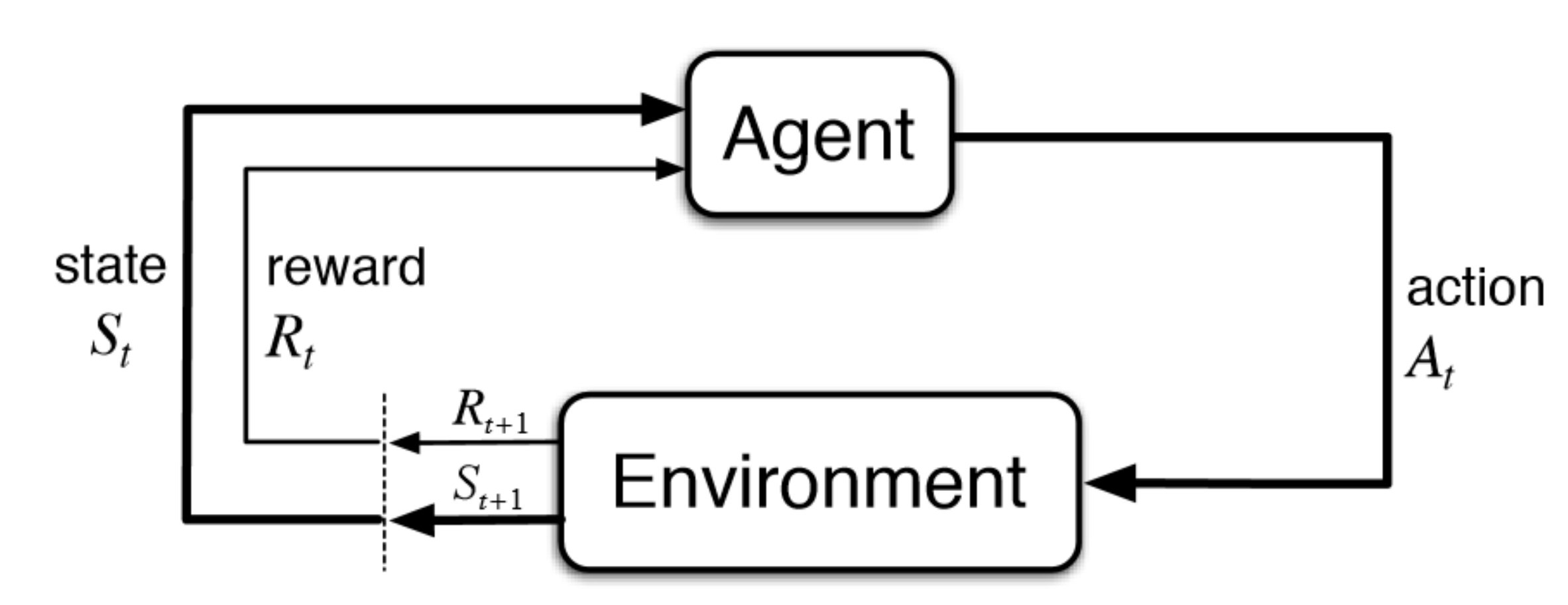}
    \caption{Components of an RL System \protect \cite{silver_introduction_nodate}}
    \label{fig:RL components}
\end{figure}

Supervised learning involves learning from a set of labeled training data. The objective is to give the learning agent the capacity to generalize responses to circumstances not in the training set \cite{silver_general_2018}. On the other hand, unsupervised learning is concerned with discovering hidden patterns and information in an unlabelled data set \cite{harmon_reinforcement_1996}. Finally, in reinforcement learning, an agent learns by interacting with an unfamiliar environment. The agent receives input from the environment in the form of a reward (or punishment) which it then utilizes to gain experience and knowledge about the environment (Figure \ref{fig:RL components}). 

The environment can be anything that processes an agent's actions and the consequences of those actions. The environment's input is an agent's action $A(t)$ performed in the current state $S(t)$, and the environment's output is the next state $S(t+1)$ and reward $R(t+1)$. A state is a location or position in each world/environment that an agent may reach or visit. The reward function $R(t)$ returns a numerical value to an agent for being in a state after performing an action. Rewards indicate whether or not a state is valuable and how valuable that state is \cite{noauthor_reinforcement_2021}. According to the reward hypothesis, all objectives may be characterized by maximizing the predicted cumulative reward. Actions are anything that an agent is permitted to do in a given context within the environment.

\begin{figure}[ht]
    \centering
    \includegraphics[scale=0.7]{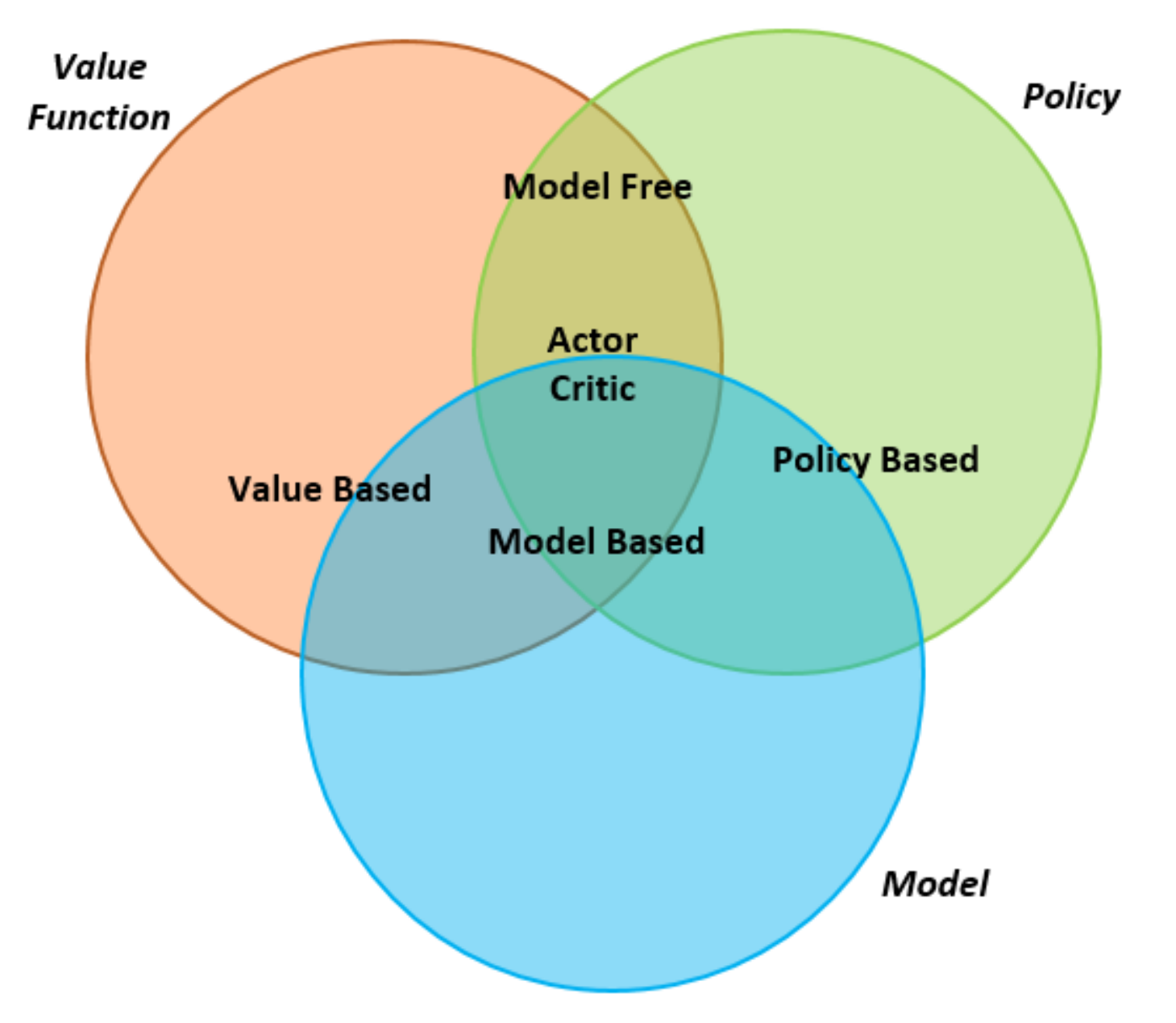}
    \caption{Taxonomy of RL Agents \protect\cite{lilianweng_long_2018}}
    \label{fig:RL approaches}
\end{figure}

An RL agent may include one or more of three components - a policy, a value function, and a model. A policy directs an agent's decision-making in each state. It is a mapping between a set of states and a set of actions. An optimum policy provides the best long-term benefits. A value function determines the quality of each state or state-action pair. A model is an agent's representation of the environment, through which the agent predicts what the environment will do next \cite{filos_reinforcement_2019, silver_general_2018}. RL agents can be classified into different classes based on which of these three components they have (Figure \ref{fig:RL approaches}). Model-free RL methods rely on trial and error to update their experience and information about the given environment because they lack knowledge of the transition model and reward function. Tables \ref{tab:model free RL methods descriptions} and \ref{tab:model free RL methods} provides a summary of some of the common model-free RL methods.

\begin{table}[p]
    \centering
    \renewcommand*{\arraystretch}{1.3}
    \caption{Table of Model-Free RL Methods.}
    \begin{tabular}{| l | l | }
    \hline
    \hline
         \textbf{Algorithm} & \textbf{Description} \\
    \hline
     A2C & Advantage Actor-Critic Algorithm \\
    \hline
         A3C & Asynchronous Advantage Actor-Critic Algorithm \\
    \hline
    DDPG & Deep Deterministic Policy Gradient \\
    \hline
    DQN & Deep Q Network \\
    \hline
     Monte Carlo & Every visit to Monte Carlo \\
    \hline
    NAF & Q-Learning with Normalized Advantage Functions \\
    \hline
    PPO & Proximal Policy Optimization \\
    \hline
    Q-learning & State-action-reward-state \\
    \hline
    Q-learning - Lambda & State action reward state with eligibility traces \\
    \hline
    REINFORCE & Monte-Carlo sampling of policy gradient methods \\
    \hline
    SAC & Soft Actor-Critic \\
    \hline
    SARSA & State-action-reward-state-action \\
    \hline
    TD3 & Twin Delayed Deep Deterministic Policy Gradient \\
    \hline
    TRPO & Trust Region Policy Optimization \\
    \hline
    \hline

    \end{tabular}
    \label{tab:model free RL methods descriptions}
\end{table}

\begin{table}[p]
    \centering
    \renewcommand*{\arraystretch}{1.3}
    \caption{Table of Model-Free RL Methods and Their Learning Mechanics.}
    \addtolength{\leftskip} {-2cm}
    \addtolength{\rightskip}{-1.5cm}
    \begin{tabular}{| l | l | l | l | l | l | }
    \hline
    \hline
    \textbf{Algorithm} & \textbf{Policy} & \textbf{Action Space} & \textbf{State Space} & \textbf{Operator} & \textbf{Class} \\
    \hline
    \hline
    A2C & On-policy & Continuous & Continuous & Advantage & Actor-Critic \\
    \hline
    A3C & On-policy & Continuous & Continuous & Advantage & Actor-Critic \\
    \hline
    DDPG & Off-policy & Continuous & Continuous & Q-value & Actor-Critic \\
    \hline
    DQN & Off-policy & Discrete & Continuous & Q-value & Value-based \\
    \hline
    Monte Carlo & Either & Discrete & Discrete & Sample means & Value-based \\
    \hline
    NAF & Off-policy & Continuous & Continuous & Advantage & Value-based \\
    \hline
    PPO & On-policy & Continuous & Continuous & Advantage & Actor-Critic \\
    \hline
    Q-learning & Off-policy & Discrete & Discrete & Q-value & Value-based \\
    \hline
    Q-learning - Lambda & Off-policy & Discrete & Discrete & Q-value & Value-based \\
    \hline
    REINFORCE & On-policy & Continuous & Continuous & Q-value & Policy-based \\
    \hline
    SAC & Off-policy & Continuous & Continuous & Advantage & Actor-Critic \\
    \hline
    SARSA & On-policy & Discrete & Discrete & Q-value & Value-based \\
    \hline
    SARSA - Lambda & On-policy & Discrete & Discrete & Q-value & Value-based \\
    \hline
    TD3 & Off-policy & Continuous & Continuous & Q-value & Actor-Critic \\
    \hline
    TRPO & On-policy & Continuous & Continuous & Advantage & Actor-Critic \\
    \hline
    \hline

    \end{tabular}
    \label{tab:model free RL methods}
\end{table}

\clearpage

\section{Reinforcement Learning Approaches}
There are mainly four common approaches to reinforcement learning - dynamic programming, Monte Carlo methods, temporal difference and policy gradient. Dynamic programming is a two-step process that uses the Bellman equation to improve the policy after the policy evaluation has been done. Dynamic programming is used when the model is fully known. Monte Carlo methods learn from raw experience episodes without modeling the environmental dynamics and compute the observed mean returns as an approximation of the predicted return. This means that learning can only occur when all episodes have been completed \cite{betancourt_deep_2021}. Temporal Difference methods learn from incomplete episodes by using bootstrapping to estimate the value. Temporal difference may be thought of as a hybrid of dynamic programming and Monte Carlo. Unlike temporal difference approaches, which rely on the value estimate to establish the optimum policy, policy gradient methods rely directly on policy estimation. The policy, $pi(a|s; theta)$, is parameterized by $theta$, and we apply gradient ascent to discover the best $theta$ producing the highest return. Policy learning can take place on- or off- policy. On-policy RL agents analyze the same policy that generated the action. Off-policy RL agents, on the other hand, analyze a policy that is not necessarily the same as the one that generated the action \cite{li_online_2014}.
\clearpage

\chapter{Trading Agents}
\section{Baseline Agents}
To properly benchmark our RL agents, we compared their performance to four baseline models. These benchmarks were chosen based on literature review and the client's recommendation. They include:
\begin{itemize}
    \item \textbf{Uniform Allocation} \newline
    This involves setting the portfolio weights such that:
    \begin{equation}
        A_{t,i}=\frac{1}{M}\ \forall\ i\ 
    \end{equation}
    
    \item \textbf{Random Allocation} \newline
    This involves setting the portfolio weights such that:
    \begin{equation}
       A_{t,i}=\frac{f\left(i\right)}{\sum_{i}^{M}{f(i)}\ }\ \newline
       where\ f\left(i\right)\ is\ a\ function\ of\ a\ random\ variable 
    \end{equation}
    
    \item \textbf{Buy And Hold} \newline
    This involves setting the portfolio weights such that:
    \begin{equation}
    A_{t,i}=c_i\ \forall\ i\ 
    0\le c_i\le\ 1, \newline
    \end{equation}
    where $c_i$ is a constant portfolio. \newline
    and $A_{t,i}$ is chosen based on the mean returns of the initial observation
    
    \item \textbf{Markowitz Model} \newline
    This involves setting the portfolio weights such that:
    \begin{equation}
        minimize_A\ \ A^T\sum A,\ \newline
    \end{equation}
            $  where\sum{is\ the\ covariance\ matrix} \newline
                subject\ to\ A^T\mu=\ \mu_{target} \newline
                and\ 1_M^TA=1 \newline
                and\ Ai\geq0\ \forall\ i \newline $
\end{itemize}

\section{Selection Criteria for RL Agents}
The RL agents were chosen using three criteria. First, the RL agent must be model-free as the focus of this work is on model-free RL. Second, the agent must have been used in similar works in literature. Finally, the agent must support continuous action and state spaces as the financial environment has continuous action and state spaces. Using table \ref{tab:model free RL methods} as the starting point, fifteen agents satisfied the first two criteria. However, only nine agents out of the fifteen agents in that table satisfied the third criterion. A3C was dropped because it is a modification of A2C that enjoys the benefits of faster computational speed when the training is distributed. Since we do not use distributed training, an A3C agent would not have offered additional advantage. Thus, at the end of the selection process, we were left with the following eight agents - A2C, DDPG, NAF, PPO, REINFORCE, SAC, TD3, TRPO.

\section{Theoretical Description of Selected RL Agents}

\subsection{Normalized Advantage Function (NAF)}

Q-learning is a temporal difference algorithm introduced in 1992 \cite{watkins1992q}. Algorithm \ref{alg:Q-learning} provides the Q-learning algorithm.

\begin{algorithm}
\caption{Q-learning \protect\cite{watkins1992q}}\label{alg:Q-learning} 
Initialize $Q(s, a)$ arbitrarily.

\For{each episode}{
    Initialize s
    
    \For{each step of episode}{
    Choose $a$ from using policy derived from Q
    
    Take action $a$ and observe reward $r$ and new state $s'$
    
    $Q(s, a)$ $\leftarrow$  $Q(s,a)$  +  $\alpha$*($r$  +  $\gamma$*$max_{a'}$*$Q(s', a')$  -  $Q(s, a)$)

    $s$ $\leftarrow$ $s'$
    
    }

   }
   
\end{algorithm}
Because Q-learning is typically computationally impractical in large action and state spaces, the Q-value function must be estimated. This may be accomplished by employing a neural network \cite{mnih2013playing}. This structure is known as a Deep Q-network (DQN), and it has served as the foundation for numerous successful applications of reinforcement learning to various tasks in recent years. DQN additionally intends to significantly enhance and stabilize the Q-learning training method through the use of two additional novel techniques:

\begin{itemize}
    \item Experience Replay
    
    Over several episodes, experience tuples comprising $(S_t, A_t, R_t, S_{t+1})$ are kept in a replay memory. During Q-learning updates, samples from the replay memory are taken at random. This increases data utilization efficiency, breaks down correlations in observation sequences, and smooths out variations in data distribution. 
    
    \item Periodically Updated Target
    
    A target network, which is just a clone of the original Q network, is established and updated on a regular basis. This improvement stabilizes the training by removing the short-term fluctuations.
\end{itemize}

However, one significant limitation of the DQN and many of its variants is that they cannot be used in continuous action spaces. \cite{gu2016continuous} proposed the normalized advantage function, a continuous variant of the Q-learning algorithm, as an alternative to policy-gradient and actor-critic approaches in 2016. The NAF model for applying Q-learning with experience replay to continuous action spaces (Algorithm \ref{alg:NAF}). \cite{gu2016continuous} also investigated the use of learned models for speeding model-free reinforcement learning and discovered that repeatedly updated local linear models are particularly successful. The authors represented the advantage function $(A)$ such that:

\begin{equation}
    Q(x, u|\theta^Q) = A(x, u|\theta^A) + V(x|\theta^U)
\end{equation}

\begin{equation}
    A(x, u|\theta^A) = -0.5(u - \mu(x|theta^\mu))^T * P(x|\theta^P) * (u - \mu(x|theta^\mu))
\end{equation}

$P(x|\theta^P)$ is a state-dependent, positive definite square matrix, which is parametrized by $L(x|\theta^P)$ which is a lower-triangular matrix whose entries come from a linear output of a neural network with the diagonal terms exponentiated.

\begin{algorithm}[H]
   \caption{Continuous Q-Learning With NAF \protect\cite{gu2016continuous}}\label{alg:NAF} 
   
   Randomly initialize normalized Q network $Q(x, u \vert \theta^Q)$.
   
   Initialize target network with weight  $\theta^{Q'} \leftarrow \theta^Q$
   
   \For{$episode = 1...$M:}{
   
   Initialize a random process $\mathcal{N}$ for action exploration
   
   Receive initial observation state $x_1 \tilde{p(x_1)}$
   
   \For{$t= 1...$T:}{
   
   Select action $u_t = \mu(x_t|\theta^\mu) + \mathcal{N}_t$
   
   Execute $u_t$ and observe $r_t$ and $x_(t+1)$
   
   Store transition $(x_t, u_t, r_t, x_(t+1) in R$
   
   \For{$iteration=1, I$}{
   
   Sample a random minibatch of m transitions from R
   
   Set $y_i = r_i + \gamma * V' * (x_(i+1)|theta^{Q'})$
   
   Update $\theta^Q $ by  minimizing $1/N\sum(y_i - Q(x_i, u_i \vert \theta^Q))^2$
   
   Update the target network: $\theta^{Q'} \leftarrow \tau*\theta^Q + (1 - \tau)*\theta^{Q'}$
   }
   }
   }
\end{algorithm}

\subsection{REINFORCE}
Policy-Gradient algorithms learn a parameterized policy that can select actions without consulting a value function. While a value function may be used to learn the policy parameters, it is not required for selecting actions. In equation \ref{PG eqn 1}, we can express the policy as the probability of action $a$ being taken at time $t$ given that the environment is in state $s$ at time $t$ with parameter $\theta$ \cite{sutton_and_barto}.

\begin{equation}
       \begin{gathered}
            \pi(a\vert s, \theta) = Pr \{ A_t = a \vert S_t = s, \theta_t = \theta \}
       \end{gathered}
    \label{PG eqn 1}
\end{equation}

Policy gradient (PG) approaches are model-free methods that attempt to maximize the RL goal without using a value function. The RL goal, also known as the performance measure $J(\theta),$ is defined as the total of rewards from the beginning state to the terminal state for an episodic task and the average return for a continuous task when policy $\pi_\theta$ is followed.
        
\begin{equation}
       \begin{gathered}
            J(\theta) \doteq v_{\pi_{\theta}}\left(s_{0}\right)=E_{\tau \sim \pi_{\theta}(\tau)}\left[\gamma^{t} r\left(s_{t}, a_{t}\right)\right]
       \end{gathered}
    \label{PG eqn 2}
\end{equation}
Where the value function $v_{\pi_\theta}(s_0)$ is the value of the expected discounted sum of rewards for a trajectory starting at state $s_0$ and following policy $\pi_\theta$ until the episode terminates. This objective can be evaluated in an unbiased manner by sampling N trajectories from the environment using policy $\pi_\theta$:

\begin{equation}
       \begin{gathered}
            J(\theta) \approx \frac{1}{N} \sum_{i=1}^{N} \sum_{t=0}^{T_{i}-1} \gamma^{t} r\left(s_{i, t}, a_{i, t}\right)
       \end{gathered}
    \label{PG eqn 3}
\end{equation}
$T_i$ is the timestep in which trajectory $\tau_i$ terminates.

The probability distribution $\pi_\theta(a \vert s)$ can be defined:
\begin{itemize}
    \item over a discrete action space, in which case the distribution is usually categorical with a softmax over the action logits.
    \item over a continuous action space, in which case the output is the parameters of a continuous distribution (e.g. the mean and variance of a gaussian).
\end{itemize}

The gradient with respect to $\theta$ according to the policy gradient theorem can be approximated over N trajectories as:

\begin{equation}
       \begin{gathered}
            \nabla_{\theta} J(\theta) \approx \frac{1}{N} \sum_{i=1}^{N}\left[\sum_{t=0}^{T_{i}-1} G_{i, t} \nabla_{\theta} \log \pi_{\theta}\left[a_{i, t} \mid s_{i, t}\right]\right]
       \end{gathered}
    \label{PG eqn 4}
\end{equation}
Where $a_{i, t}$ is the action taken at time $t$ of episode $i$ at state $s_{i,t}$, $T_i$ is the timestep in which trajectory $\tau_i$ terminates and $G_{i,t}$ is a function of the reward assigned to this action

For REINFORCE, $G_{i, t}$ is the sum of rewards in trajectory $i$
\begin{equation}
       \begin{gathered}
            G_{i, t}=\sum_{t^{\prime}=0}^{T_{i}-1} r\left(s_{i, t^{\prime}}, a_{i, t^{\prime}}\right)
       \end{gathered}
    \label{PG eqn 5}
\end{equation}

\begin{algorithm}[H]
    \caption{REINFORCE: Monte-Carlo Policy Gradient Control (episodic)}\label{alg:reinforce}
    Initialize policy network with weights $\theta$

    \For{each episode $\{s_0, a_0, r_2\ ...\ s_{T-1}, a_{T-1}, r_T\}$ sampled from policy $\pi_\theta$:}{

        \For{$t = 0...T-1$:}{
        
            Evaluate the gradient
            
            \hspace{50pt} $\nabla_{\theta} J(\theta) \approx \frac{1}{N} \sum_{i=1}^{N}\left[\sum_{t=0}^{T_{i}-1} G_{i, t} \nabla_{\theta} \log \pi_{\theta}\left[a_{i, t} \mid s_{i, t}\right]\right]$

            Update the policy parameters
            
            \hspace{50pt} $\theta \leftarrow \theta + \alpha \triangledown_\theta J(\theta)$ \\
        }
    }
\end{algorithm}

\textbf{Limitations}
\begin{enumerate}
\item The procedure of updating is inefficient. The trajectory is deleted after performing the policy and changing the parameters. 
\item The gradient estimate is noisy, and there is a chance that the gathered trajectory does not accurately represent the policy item. 
\item There is no apparent credit assignment. A trajectory can contain numerous good or harmful activities, and whether or not these behaviours are reinforced is only determined by the ultimate product.
\end{enumerate}

Other Policy Gradient methods like A2C, DDPG, TD3, SAC, and PPO were created to overcome the limitations of REINFORCE.

\subsection{Deep Deterministic Policy Gradient (DDPG)}
Following the success of the Deep-Q Learning algorithm, which beat humans in Atari games, DeepMind applied the same concept to physics challenges, where the action space is considerably larger than in Atari games. Deep Q-Learning performed well in high-dimensional state spaces but not in high-dimensional action spaces (continuous action). To cope with high-dimensional (continuous) action spaces, DDPG blends Deep Learning and Reinforcement Learning approaches. DDPG employs the concepts of an experience replay buffer, in which the network is trained off-policy by sampling experience batches, and target networks, in which copies of the network are created for use in objective functions to avoid divergence and instability in complex and non-linear function approximators such as neural networks \cite{lillicrap_continuous_2019}.

Aside from using a neural network to parameterize the Q-function "critic," as shown in DQN, we also have the policy network called "actor" to parameterize the policy function. The policy is simply the behaviour of the agent, "a mapping from state to action" in the case of a deterministic policy or "a distribution of actions" in the case of a stochastic policy. Since we have two networks, there are two sets of parameters to update:
\begin{enumerate}
    \item The parameters of the policy network have to be updated in order to maximize the performance measure $J$ defined in the policy gradient theorem 
    \item The parameters of the critic network are updated in order to minimize the temporal difference loss $L$
\end{enumerate}

\begin{equation}
       L(w) = \frac{1}{N} \sum_{i} (y_i - \hat{q}(s_i, a_i, w))^2
    \label{DDPG Loss}
\end{equation}

\begin{equation}
       \triangledown_{\theta}J(\theta) \approx \frac{1}{N} \sum_{i} \triangledown_{a}\hat{q}(s,a,w)\vert_{s=S_{i}, a=\pi(S_{i})} \triangledown_\theta \pi(s, \theta) \vert_{s=S_{i}}
    \label{DDPG Performance Measure}
\end{equation}

To maximize the Q-value function while reducing the temporal difference loss, we must enhance the performance measure J. The actor takes the state as input and outputs an action, whereas the critic takes both the state and the action as input and outputs the value of the Q function. The critic uses gradient temporal-difference learning, whilst the actor parameters are discovered using the Policy gradient theorem. The essential principle of this design is that the policy network acts, resulting in an action, and the Q-network critiques that action.

The use of non-linear function approximators such as neural networks, which are required to generalize on vast state spaces, means that convergence is no longer assured, as it was with Q learning. As a result, experience replay is required in order to generate independent and identically dispersed samples. In addition, target networks must be used to avoid divergence when upgrading the critic network. In DDPG, parameters are changed differently than in DQN, where the target network is updated every C steps. Following the "soft" update, the parameters of the target networks are changed at each time step as shown:

\begin{equation}
       \begin{gathered}
            w^{-} \leftarrow \tau w + (1 - \tau)w^{-} \\
       \theta^{-} \leftarrow \tau \theta + (1 - \tau)\theta^{-} 
       \end{gathered}
    \label{DDPG Update step}
\end{equation}

where $\tau \ll 1$, $w^- =$ weights of target $Q$ network, $\theta^- =$ weights of target policy network
($\ll =$ much less than)

The weights of target networks are limited to fluctuate slowly using "soft" updates, boosting the stability of learning and convergence outcomes. The target network is then utilized instead of the Q-network in the temporal difference loss. In algorithms such as DDPG, the problem of exploration may be addressed quite easily and independently of the learning process. The actor policy is then supplemented with noise taken from a noise process N to create the exploration policy. The exploration policy then becomes:

\begin{equation}
            \pi'(S_t) = \pi(S_t, \theta) + v \\
    \label{DDPG Policy Update}
\end{equation}

Where $v$ is an Ornstein-Uhlenbeck process - a stochastic method capable of producing temporally coordinated actions that provide smooth exploration in physical control issues.

\begin{algorithm}[H]
    \caption{DDPG \protect\cite{lillicrap_continuous_2019}}\label{alg:ddpg}
    Randomly initialize critic network $Q(s,a \vert \theta^Q)$ and actor $\mu(s \vert \theta^{\mu})$ with weights $\theta^Q$ and $\theta^\mu$
    
    Initialize target network $Q'$ and $\mu'$ with weights $\theta^{Q'} \leftarrow \theta^Q, \theta^{\mu'} \leftarrow \theta^\mu$
    
    Initialize replay buffer R
    
    \For{$episode = 1...$M:}{
        Initialize a random process $\mathcal{N}$ for action exploration
        
        Receive initial observation state $s_1$
        
        \For{$t = 1...$T:}{
            Select action $a_t = \mu(s_t \vert \theta^\mu) + \mathcal{N}_t$ according to the current policy and exploration noise
            
            Execute action $a_t$ and observe reward $r_t$ and observe new state $s_{t + 1}$
            
            Store transition $(s_t, a_t, r_t, s_{t+1})$ in $R$
            
            Sample a random minibatch of $N$ transitions $(s_i, a_i, r_i, s_{i+1})$ from $R$
            
            Set $y_i = r_i + \gamma Q'(s_{i+1}, \mu'(s_{i+1} \vert \theta^{\mu'}) \vert \theta^{Q'})$
            
            Update critic by minimizing the loss: $L(w) = \frac{1}{N} \sum_{i} (y_i - \hat{q}(s_i, a_i, w))^2$
            
            Update the the actor policy using the sampled policy gradient:

            \hspace{25pt} $\triangledown_{\theta^\mu}J(\theta) \approx \frac{1}{N} \sum_{i} \triangledown_{a}Q(s,a \vert \theta^Q)\vert_{s=s_{i}, a=\mu(s_{i})} \triangledown_{\theta^\mu} \mu(s \vert \theta^\mu) \vert_{s_{i}}$
            
            Update the target networks:
            
            \hspace{50pt} $\theta^{Q'} \leftarrow \tau \theta^{Q} + (1 - \tau)\theta^{Q'}$ \\
            \hspace{50pt} $\theta^{\mu'} \leftarrow \tau \theta^{\mu} + (1 - \tau)\theta^{\mu'}$
            
        }
    }
\end{algorithm}

\subsection{Twin Delayed Deep Deterministic Policy Gradient (TD3)}
The DQN method is known to exhibit overestimation bias, which means that it overestimates the value function. This is due to the fact that the goal Q value is an approximation, and choosing the maximum over an estimate implies we are strengthening the approximation inaccuracy. This difficulty prompted various enhancements to the underlying DQN algorithm. TD3 uses numerous algorithmic methods on DDPG, a network designed to improve on the DQN, to limit the possibility of overestimation bias drastically. The algorithmic methods are clipped double Q-Learning, delayed policy/targets updates, and target policy smoothing.

TD3 employs six neural networks: one actor, two critics, and the target networks that correspond to them. Clipped Double Q-Learning employs the least estimation between the two actor reviewers in order to favor underestimating the value function, which is difficult to transmit through the training process. To limit the volatility in the value estimation, TD3 updates the policy at a reduced frequency (Delayed Policy and Targets Updates). 

The policy network remains unchanged until the value error is small enough. Furthermore, rather than just duplicating the weights after k steps, the target networks are updated using the Polyak averaging approach. Finally, to avoid overfitting, TD3 smooths the value function by adding a little amount of clipped random noises to the chosen action and averaging over mini-batches. Algorithm \ref{alg:TD3} depicts the TD3 framework and the places where these algorithmic tricks were used.

\begin{algorithm}[H]
    \caption{TD3 \protect\cite{fujimoto2018addressing}}\label{alg:TD3} 
    Initialize critic networks $Q_{\theta_1}$, $Q_{\theta_2}$ and actor network $\pi_{\phi}$ with random parameters $\theta_1$, $\theta_2$, $\phi$
    
    Initialize target networks $\theta'_1 \leftarrow \theta_1$, $\theta'_2 \leftarrow \theta_2$, $\phi' \leftarrow \phi$
    
    Initialize replay buffer $\mathcal{B}$
    
    \For{$t = 1...$T:}{
        Select action with exploration noise $a \sim \pi(s) + \epsilon$, $\epsilon \sim \mathcal{N}(0, \sigma)$ and observe reward $r$ and new state $s'$
        
        Store transition tuple $(s,a,r,s')$ in $\mathcal{B}$
        
        Sample mini-batch of $N$ transitions $(s,a,r,s')$ from $\mathcal{B}$
        
        $\Tilde{a} \leftarrow \pi_{\phi'}(s) + \epsilon$, \hspace{15pt} $\epsilon \sim clip(\mathcal{N}(0, \Tilde{\sigma}), -c,c)$ 
        \tcp*{Target Policy Smoothing}
        
        $y \leftarrow r + \gamma\min_{i=1,2}Q_{\theta'_i}(s', \Tilde{a})$
        \tcp*{Clipped Double Q-learning}
        
        Update critics $\theta_i \leftarrow min_{\theta_i}N^{-1}\sum(y - Q_{\theta_i}(s,a))^2$
        
        \If{t \textbf{mod} d}{ \tcc{Delayed update of target and policy networks}
            Update $\phi$ by the deterministic policy gradient:
            
            $\triangledown_{\phi}J(\phi) = N^{-1} \sum \triangledown_{a}Q_{\theta_1}(s,a)\vert_{a=\pi_{\phi}(s)} \triangledown_{\phi} \pi_{\phi}(s)$
            
            Update the target networks:
            
            \hspace{50pt} $\theta'_{i} \leftarrow \tau \theta_{i} + (1 - \tau)\theta'_{i}$ \\
            \hspace{50pt} $\phi'_{i} \leftarrow \tau \phi + (1 - \tau)\phi'$
         }
         
    }
\end{algorithm}

\subsection{Advantage Actor Critic (A2C)}
A2C is a policy gradient method that combines two types of reinforcement learning algorithms: policy-based and value-based. The actor-critic algorithm is composed of two distinct structures: one for storing and updating the value function and the other for storing the updated policy. The agent chooses the action based on the policy rather than the value function, where the policy component is called the \textbf{actor}, which conducts an action, changes the value of the function, and uses the value function to assess the action, and the value function part is called the \textbf{critic}. To lower the variance of the policy gradient, it employs an \textbf{advantage} (equation \ref{a2c advantage function}). The critic network assesses the advantage function rather than the value function only \cite{tang_actor-critic-based_2018}. 

\begin{equation}
        \triangledown_\theta J(\theta) \sim \sum\limits_{t=0}^{T-1}
        \triangledown_\theta log\pi_\theta(a_t \vert s_t)A(s_t, a_t)
        \label{a2c advantage function} 
\end{equation}

Thus, the evaluation of an action examines not only how excellent the action is, but also how the action may be improved so that the high variance of the policy networks is lowered and the model becomes more resilient \cite{konda_actor-critic_2000}. The value function of the critic component can also make use of temporal difference error (TD error) computed using the TD learning approach.

\begin{figure}[h!]
    \centering
    \includegraphics[scale=0.1]{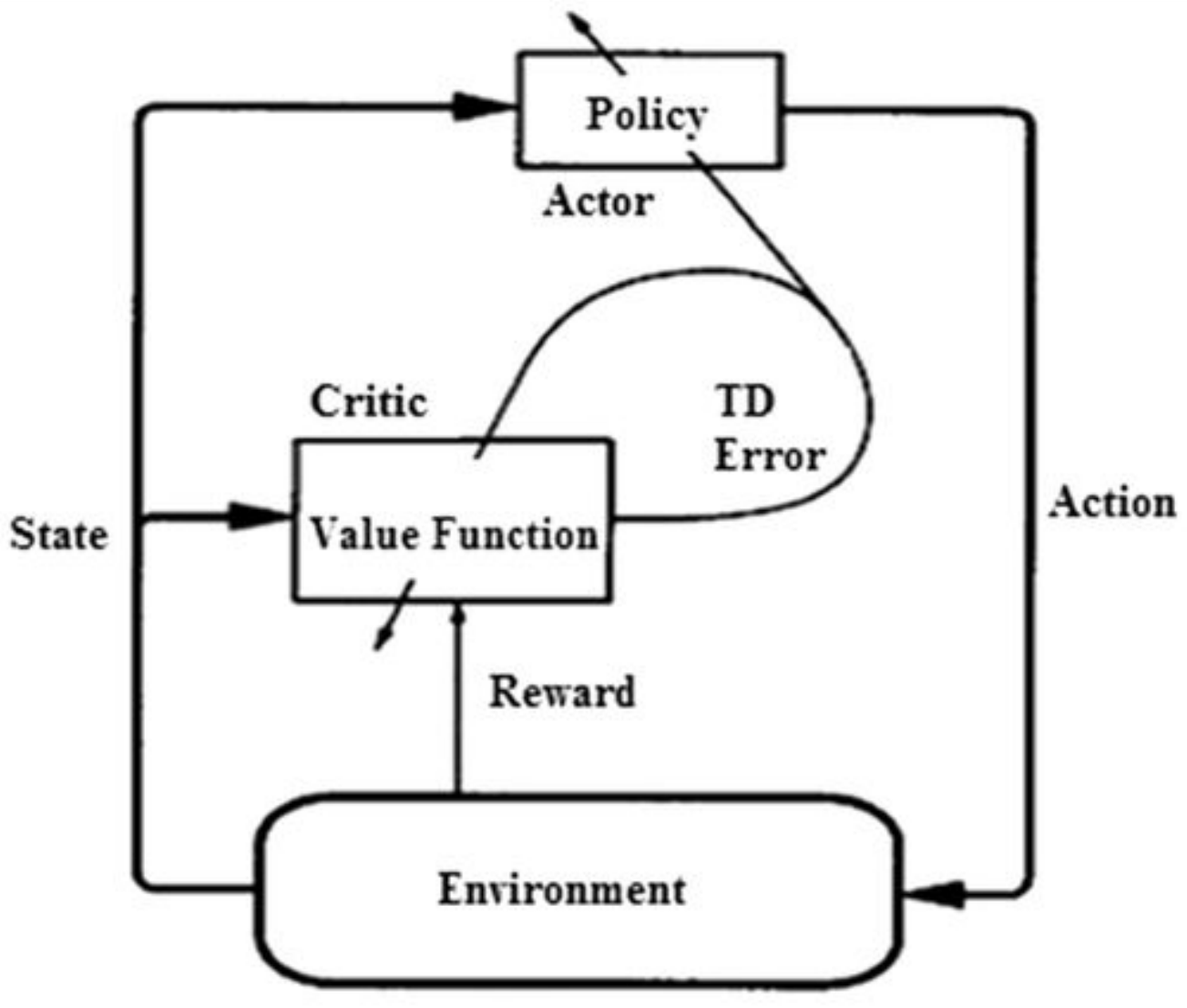}
    \caption{Actor Critic Algorithm Framework \protect\cite{sutton_and_barto}}
    \label{fig:Actor Critic Framework}
\end{figure}

The advantage of the actor-critic algorithm is that it separates the policy from the value function by learning the value function and the policy function using linear approximation, where the critic part is the value function approximator, learning the estimate function, and then passing to the actor part. The actor is a policy approximator that learns a random strategy and selects an action using the gradient-based policy updating approach (Fig. \ref{fig:Actor Critic Framework}).

The Policy gradient is defined as follows:

\begin{equation}
        \triangledown_\theta J(\theta) = \mathbb{E}_{\tau} \left[ \sum\limits_{t=0}^{T-1}
        \triangledown_\theta log\pi_\theta(a_t \vert s_t)G_t \right]
        \label{a2c policy gradient} 
\end{equation}

Introducing baseline $b(s)$:

\begin{equation}
        \triangledown_\theta J(\theta) = \mathbb{E} \left[ \sum\limits_{t=0}^{T-1}
        \triangledown_\theta log\pi_\theta(a_t \vert s_t)(G_t - b(s_t)) \right]
        \label{a2c policy gradient with baseline} 
\end{equation}

\begin{equation}
        \mathbb{E}_{r_{t+1},s_{t+1},...,r_T,s_T}[G_t] = Q(s_t, a_t)
        \label{a2c policy gradient with baseline 3} 
\end{equation}

Plugging that in, we can rewrite the update equation as such:

\begin{equation}
        \triangledown_\theta J(\theta) = \mathbb{E}_{s_0, a_0, ..., s_t, a_t} \left[ \sum\limits_{t=0}^{T-1}
        \triangledown_\theta log\pi_\theta(a_t \vert s_t) \right] Q_w(s_t, a_t)
        \label{a2c update equation} 
\end{equation}

\begin{equation}
        = \mathbb{E}_{\tau} \left[ \sum\limits_{t=0}^{T-1}
        \triangledown_\theta log\pi_\theta(a_t \vert s_t) Q_w(s_t, a_t) \right]
        \label{a2c update equation 2} 
\end{equation}

\begin{algorithm}[H]
    \caption{Actor Critic}\label{alg:a2c}
    Initialize parameters $s,\theta, w$ and learning rates $\alpha_\theta, \alpha_w$ sample $a \sim \pi_\theta(a \vert s)$
    
    \For{$t\gets1, 2, ...$T:}{
        Sample reward $r_t \sim R(s,a)$ and next state $s' \sim P(s'\vert s,a)$
        
        Then sample the next action $a' \sim \pi_\theta(a' \vert s')$
        
        Update the policy parameters: $\theta \leftarrow \theta + \alpha_{\theta}Q_{w}(s,a)\triangledown_{\theta}log\pi_{\theta}(a \vert s)$
        
        Compute the correction (TD error) for action-value at time t:
        
        \hspace{25pt} $\delta_{t} = r_t + \gamma Q_{w}(s',a') - Q_{w}(s,a) $
        
        and use it to update the parameters of Q function:
        
        \hspace{25pt} $w \leftarrow w + \alpha_{w}\delta_{t}\triangledown_{w}Q_{w}(s,a) $
        
        Move to a $\leftarrow a'$ and $s \leftarrow s'$
        
    }
\end{algorithm} 

\subsection{Soft Actor Critic(SAC)}
As a bridge between stochastic policy optimization and DDPG techniques, the Soft Actor-Critic Algorithm is an off-policy algorithm that optimizes a stochastic policy. Entropy regularization is the primary aspect of SAC. The policy is trained to optimize a trade-off between anticipated return and entropy, which is a measure of the policy's unpredictability. SAC is thoroughly described by first establishing the entropy regularized reinforcement learning setup and the value functions associated with it \cite{haarnoja2018sac}.

The SAC algorithm learns a policy as well as two Q- functions, Q1 and Q2. There are two basic SAC variants: one that utilizes a constant entropy regularization coefficient and another that enforces an entropy restriction by altering the training process. The constant entropy regularization coefficient is employed for ease in spinning up, although the entropy-constrained variation is more widely utilized, according to \cite{haarnoja2018sac}. The Q-functions are taught in a manner similar to the TD3 mentioned in this study, with a few major modifications.

In each stage, the policy should be implemented in order to maximize projected future returns and expected future entropy. It should aim to maximize  $V^{\pi}(s)$, which we expand out into
$$
\begin{aligned}
V^{\pi}(s) &=\underset{a \sim \pi}{\mathrm{E}}\left[Q^{\pi}(s, a)\right]+\alpha H(\pi(\cdot \mid s)) \\
&=\underset{a \sim \pi}{\mathrm{E}}\left[Q^{\pi}(s, a)-\alpha \log \pi(a \mid s)\right]
\end{aligned}
$$

\begin{algorithm}[H]
    \caption{Soft Actor Critic:}\label{alg:SAC}
    \SetKwInOut{Input}{Input}
    \Input {initial policy parameters $\theta$, Q-function parameters $\phi_{1}, \phi_{2}$, empty replay buffer $\mathcal{D}$}
    {Set target parameters equal to main parameters $\phi_{\text {targ }, 1} \leftarrow \phi_{1}, \phi_{\text {targ }, 2} \leftarrow \phi_{2}$} 
    Repeat the following steps until convergence:\\
    Observe state $s$ and select action $a \sim \pi_{\theta}(\cdot \mid s)$\\
    Execute $a$ in the environment\\
    Observe next state $s^{\prime}$, reward $r$, and done signal $d$ to indicate whether $s^{\prime}$ is terminal Store $\left(s, a, r, s^{\prime}, d\right)$ in replay buffer $\mathcal{D}$\\
    If $s^{\prime}$ is terminal, reset environment state.\\
    if it's time to update then\\
    
    \For{$k = 0, 1, 2, ...$}{
        Randomly sample a batch of transitions, $B=\left\{\left(s, a, r, s^{\prime}, d\right)\right\}$ from $\mathcal{D}$ Compute targets for the Q functions:
        $$
        y\left(r, s^{\prime}, d\right)=r+\gamma(1-d)\left(\min _{i=1,2} Q_{\phi_{\text {targ }, i}}\left(s^{\prime}, \tilde{a}^{\prime}\right)-\alpha \log \pi_{\theta}\left(\tilde{a}^{\prime} \mid s^{\prime}\right)\right), \quad \tilde{a}^{\prime} \sim \pi_{\theta}\left(\cdot \mid s^{\prime}\right)
        $$
        Update Q-functions by one step of gradient descent using
        $$
        \nabla_{\phi_{i}} \frac{1}{|B|} \sum_{\left(s, a, r, s^{\prime}, d\right) \in B}\left(Q_{\phi_{i}}(s, a)-y\left(r, s^{\prime}, d\right)\right)^{2} \quad \text { for } i=1,2
        $$
        Update policy by one step of gradient ascent using
        $$
        \nabla_{\theta} \frac{1}{|B|} \sum_{s \in B}\left(\min _{i=1,2} Q_{\phi_{i}}\left(s, \tilde{a}_{\theta}(s)\right)-\alpha \log \pi_{\theta}\left(\tilde{a}_{\theta}(s) \mid s\right)\right)
        $$
        where $\tilde{a}_{\theta}(s)$ is a sample from $\pi_{\theta}(\cdot \mid s)$ which is differentiable wrt $\theta$ via the reparametrization trick.
        Update target networks with
        $$
        \phi_{\mathrm{targ}, i} \leftarrow \rho \phi_{\mathrm{targ}, i}+(1-\rho) \phi_{i} \quad \text { for } i=1,2
        $$
    } 
\end{algorithm} 

\subsection{Trust Region Policy Optimization (TRPO)}
TRPO is based on trust region optimization, which ensures monotonic improvement by adding trust region restrictions to satisfy how near the new and old policies may be \cite{Schulman2015-kg}. The restriction is defined in terms of KL divergence, which is a measure of the distance between probability distributions \cite{Joyce2011}.

In TRPO, the goal is to optimize the objective function which is denoted as 
$\eta(\pi)$. Consider an infinite-horizon discounted Markov decision process (MDP), defined by the tuple $\left(\mathcal{S}, \mathcal{A}, P, r, \rho_{0}, \gamma\right)
$, where $\mathcal{S}$ is a finite set of states, $\mathcal{A}$ is a finite set of actions, $
P: \mathcal{S} \times \mathcal{A} \times \mathcal{S} \rightarrow \mathbb{R}
$ is the transition probability distribution, $r: \mathcal{S} \rightarrow \mathbb{R}$ is the reward function, $\rho_{0}: \mathcal{S} \rightarrow \mathbb{R}
$ is the distribution of the initial state $
s_{0}
$, and $
\gamma \in(0,1)
$ is the discount factor \cite{Schulman2015-kg}.

Let $\pi$ denote a stochastic policy $\pi: \mathcal{S} \times \mathcal{A} \rightarrow[0,1]$, and let $\eta(\pi)$ denote its expected discounted reward:
$$
\begin{aligned}
&\eta(\pi)=\mathbb{E}_{s_{0}, a_{0}, \ldots}\left[\sum_{t=0}^{\infty} \gamma^{t} r\left(s_{t}\right)\right], \text { where } \\
&s_{0} \sim \rho_{0}\left(s_{0}\right), a_{t} \sim \pi\left(a_{t} \mid s_{t}\right), s_{t+1} \sim P\left(s_{t+1} \mid s_{t}, a_{t}\right)
\end{aligned}
$$
In TRPO, the state-action value function $Q_{\pi}$, the value function $V_{\pi}$, and the advantage function $A_{\pi}$ are defined as:
$$
\begin{aligned}
&Q_{\pi}\left(s_{t}, a_{t}\right)=\mathbb{E}_{s_{t+1}, a_{t+1}, \ldots}\left[\sum_{l=0}^{\infty} \gamma^{l} r\left(s_{t+l}\right)\right] \\
&V_{\pi}\left(s_{t}\right)=\mathbb{E}_{a_{t}, s_{t+1}, \ldots}\left[\sum_{l=0}^{\infty} \gamma^{l} r\left(s_{t+l}\right)\right] \\
&A_{\pi}(s, a)=Q_{\pi}(s, a)-V_{\pi}(s), \text { where } \\
&a_{t} \sim \pi\left(a_{t} \mid s_{t}\right), s_{t+1} \sim P\left(s_{t+1} \mid s_{t}, a_{t}\right) \text { for } t \geq 0
\end{aligned}
$$

The following useful identity expresses the expected return of another policy $\tilde{\pi}$ in terms of the advantage over $\pi$, accumulated over time-steps \cite{Schulman2015-kg}:

\begin{equation}
    \eta(\tilde{\pi})=\eta(\pi)+\mathbb{E}_{s_{0}, a_{0}, \cdots \sim \bar{\pi}}\left[\sum_{t=0}^{\infty} \gamma^{t} A_{\pi}\left(s_{t}, a_{t}\right)\right]
    \label{TRPO eqn 1}
\end{equation}

where the notation $\mathbb{E}_{s_{0}, a_{0}, \ldots \sim \bar{\pi}}[\ldots]$ indicates that actions are sampled $a_{t} \sim \tilde{\pi}\left(\cdot \mid s_{t}\right) .$ Let $\rho_{\pi}$ be the (unnormalized) discounted visitation frequencies
$$
\rho_{\pi}(s)=P\left(s_{0}=s\right)+\gamma P\left(s_{1}=s\right)+\gamma^{2} P\left(s_{2}=s\right)+\ldots
$$
where $s_{0} \sim \rho_{0}$ and the actions are chosen according to $\pi$. \ref{TRPO eqn 1} can be re-written with a sum over states instead of time steps:

\begin{equation}
    \begin{aligned}
        \eta(\tilde{\pi}) &=\eta(\pi)+\sum_{t=0}^{\infty} \sum_{s} P\left(s_{t}=s \mid \tilde{\pi}\right) \sum_{a} \tilde{\pi}(a \mid s) \gamma^{t} A_{\pi}(s, a) \\
        &=\eta(\pi)+\sum_{s} \sum_{t=0}^{\infty} \gamma^{t} P\left(s_{t}=s \mid \tilde{\pi}\right) \sum_{a} \tilde{\pi}(a \mid s) A_{\pi}(s, a) \\
        &=\eta(\pi)+\sum_{s} \rho_{\bar{\pi}}(s) \sum_{a} \tilde{\pi}(a \mid s) A_{\pi}(s, a)
    \end{aligned}
    \label{TRPO eqn 2}
\end{equation}

Equation \ref{TRPO eqn 2} implies that any policy update $\pi \rightarrow \tilde{\pi}$ that has a non-negative expected advantage at every state $s$, i.e., $\sum_{a} \tilde{\pi}(a \mid s) A_{\pi}(s, a) \geq 0$, is guaranteed to increase the policy performance $\eta$, or leave it constant in the case that the expected advantage is zero everywhere. This implies the classic result that the update performed by exact policy iteration, which uses the deterministic policy $\bar{\pi}(s)=\arg \max _{a} A_{\pi}(s, a)$, improves the policy if there is at least one state-action pair with a positive advantage value and nonzero state visitation probability; otherwise, the algorithm has converged to the optimal policy. However, in the approximate setting, it will typically be unavoidable, due to estimation and approximation error, that there will be some states $s$ for which the expected advantage is negative, that is, $\sum_{a} \tilde{\pi}(a \mid s) A_{\pi}(s, a)<0$. The complex dependency of $\rho_{\bar{\pi}}(s)$ on $\tilde{\pi}$ makes Equation \ref{TRPO eqn 2} difficult to optimize directly. Instead, the following local approximation to $\eta$ is introduced \cite{Schulman2015-kg}:

\begin{equation}
    L_{\pi}(\tilde{\pi})=\eta(\pi)+\sum_{s} \rho_{\pi}(s) \sum_{a} \tilde{\pi}(a \mid s) A_{\pi}(s, a)
    \label{TRPO eqn 3}
\end{equation}

$L_{\pi}$ uses the visitation frequency $\rho_{\pi}$ rather than $\rho_{\bar{\pi}}$, ignoring changes in state visitation density due to changes in the policy. However, when using a parameterized policy $\pi_{\theta}$, where $\pi_{\theta}(a \mid s)$ is a differentiable function of the parameter vector $\theta$, then $L_{\pi}$ matches $\eta$ to first order. That is, for any parameter value $\theta_{0}$,

\begin{equation}
    \begin{aligned}
        L_{\pi_{\theta}}\left(\pi_{\theta_{0}}\right) &=\eta\left(\pi_{\theta_{0}}\right) \\
        \left.\nabla_{\theta} L_{\pi_{\theta_{0}}}\left(\pi_{\theta}\right)\right|_{\theta=\theta_{0}} &=\left.\nabla_{\theta} \eta\left(\pi_{\theta}\right)\right|_{\theta=\theta_{0}}
    \end{aligned}
    \label{TRPO eqn 4}
\end{equation}

Equation \ref{TRPO eqn 3} implies that a sufficiently small step $\pi_{\theta_{0}} \rightarrow \bar{\pi}$ that improves $L_{\pi_{\theta_{\text {old }}}}$ will also improve $\eta$, but does not give us any guidance on how big of a step to take \cite{Schulman2015-kg}.

To address this issue, \cite{inproceedings} proposed a policy updating scheme called conservative policy iteration, for which they could provide explicit lower bounds on the improvement of $\eta$. To define the conservative policy iteration update, let $\pi_{\text {old }}$ denote the current policy, and let $\pi^{\prime}=\arg \max _{\pi^{\prime}} L_{\pi_{\text {old }}}\left(\pi^{\prime}\right) .$ The new policy $\pi_{\text {new }}$ was defined to be the following mixture \cite{Schulman2015-kg}:

\begin{equation}
    \pi_{\text {new }}(a \mid s)=(1-\alpha) \pi_{\text {old }}(a \mid s)+\alpha \pi^{\prime}(a \mid s)
    \label{TRPO eqn 6}
\end{equation}

\cite{inproceedings} derived the following lower bound:

\begin{equation}
    \begin{aligned}
    \eta\left(\pi_{\text {new }}\right) & \geq L_{\pi_{\text {old }}}\left(\pi_{\text {new }}\right)-\frac{2 \epsilon \gamma}{(1-\gamma)^{2}} \alpha^{2} \\
    & \text { where } \epsilon=\max _{s}\left|\mathbb{E}_{a \sim \pi^{\prime}(a \mid s)}\left[A_{\pi}(s, a)\right]\right|
    \end{aligned}
    \label{TRPO eqn 7}
\end{equation}

Equation \ref{TRPO eqn 7}, which applies to conservative policy iteration, implies that a policy update that improves the right hand side is guaranteed to improve the true performance $\eta$. Our principal theoretical result is that the policy improvement bound in Equation \ref{TRPO eqn 7} can be extended to general stochastic policies, rather than just mixture policies, by replacing $\alpha$ with a distance measure between $\pi$ and $\tilde{\pi}$, and changing the constant $\epsilon$ appropriately. Since mixture policies are rarely used in practice, this result is crucial for extending the improvement guarantee to practical problems. The particular distance measure we use is the total variation divergence, which is defined by $D_{T V}(p \| q)=\frac{1}{2} \sum_{i}\left|p_{i}-q_{i}\right|$ for discrete probability distributions $p, q .^{1}$ Define $D_{\mathrm{TV}}^{\max }(\pi, \tilde{\pi})$ as

\begin{equation}
D_{\mathrm{TV}}^{\max }(\pi, \tilde{\pi})=\max _{s} D_{T V}(\pi(\cdot \mid s) \| \tilde{\pi}(\cdot \mid s))
\label{TRPO eqn 8}
\end{equation}

Trust region policy optimization uses a constraint on the KL divergence rather than a penalty to robustly allow large updates. Thus, by performing the following maximization, we are guaranteed to improve the true objective $\eta$ \cite{Schulman2015-kg}:

\begin{equation}
\underset{\theta}{\operatorname{maximize}}\left[L_{\theta_{\text {old }}}(\theta)-C D_{\mathrm{KL}}^{\max }\left(\theta_{\text {old }}, \theta\right)\right]
\label{TRPO eqn 9}
\end{equation}

Therefore, TRPO solves the following optimization problem to generate a policy update \cite{Schulman2015-kg}:
\begin{equation}
\begin{aligned}
&\underset{\theta}{\operatorname{maximize}} L_{\theta_{\text {old }}}(\theta) \\
&\text { subject to } \bar{D}_{\mathrm{KL}}^{\rho_{\text {old }}}\left(\theta_{\text {old }}, \theta\right) \leq \delta
\end{aligned}
\label{TRPO eqn 10}
\end{equation}

TRPO seeks to solve the following optimization problem, obtained by expanding $L_{\theta_{\text {old }}}$ in Equation (2.10):
\begin{equation}
\begin{array}{r}
\underset{\theta}{\operatorname{maximize}} \sum_{s} \rho_{\theta_{\text {old }}}(s) \sum_{a} \pi_{\theta}(a \mid s) A_{\theta_{\text {old }}}(s, a) \\
\text { subject to } \bar{D}_{\mathrm{KL}}^{\rho_{\theta}}\left(\theta_{\text {old }}, \theta\right) \leq \delta
\end{array}
\label{TRPO eqn 11}
\end{equation}

The optimization problem in Equation \ref{TRPO eqn 11} is exactly equivalent to the following one, written in terms of expectations \cite{Schulman2015-kg}:

\begin{equation}
\begin{aligned}
&\underset{\theta}{\operatorname{maximize}} \mathbb{E}_{s \sim \rho_{\theta_{\text {old }}}, a \sim q}\left[\frac{\pi_{\theta}(a \mid s)}{q(a \mid s)} Q_{\theta_{\text {old }}}(s, a)\right] \\
&\text { subject to } \mathbb{E}_{s \sim \rho_{\theta_{\text {old }}}}\left[D_{\mathrm{KL}}\left(\pi_{\theta_{\text {old }}}(\cdot \mid s) \| \pi_{\theta}(\cdot \mid s)\right)\right.
\end{aligned}
\label{TRPO eqn 12}
\end{equation}

\begin{algorithm}[H]
    \caption{Trust Region Policy Optimization
    \protect\cite{Schulman2015-kg}}\label{alg:trpo}
    \SetKwInOut{Input}{Input}
    \Input{initial policy parameters $\theta_{0}$, initial value function         paramaters $\phi_{0}$}
    \SetKwInOut{Hyperparameters}{Hyperparameters}
    \Hyperparameters{KL-divergence limit $\delta$, backtracking coefficient $\alpha$, maximum number of backtracking steps $K$}
    \For{$k = 0, 1, 2, ...$}{
        Collect set of trajectories $\mathcal{D}_{k}=\left\{\tau_{i}\right\}$ by running policy $\pi_{k}=\pi\left(\theta_{k}\right)$ in the environment. Compute rewards-to-go $\hat{R}_{t} .$
        Compute advantage estimates, $\hat{A}_{t}$ (using any method of advantage estimation) based on the current value function $V_{\phi_{k}}$.
        Estimate policy gradient as
        $$
        \hat{g}_{k}=\left.\frac{1}{\left|\mathcal{D}_{k}\right|} \sum_{\tau \in \mathcal{D}_{k}} \sum_{t=0}^{T} \nabla_{\theta} \log \pi_{\theta}\left(a_{t} \mid s_{t}\right)\right|_{\theta_{k}} \hat{A}_{t}
        $$
        Use the conjugate gradient algorithm to compute
        $$
        \hat{x}_{k} \approx \hat{H}_{k}^{-1} \hat{g}_{k}
        $$
        where $\hat{H}_{k}$ is the Hessian of the sample average KL-divergence. Update the policy by backtracking line search with
        $$
        \theta_{k+1}=\theta_{k}+\alpha^{j} \sqrt{\frac{2 \delta}{\hat{x}_{k}^{T} \hat{H}_{k} \hat{x}_{k}}} \hat{x}_{k}
        $$
        where $j \in\{0,1,2, \ldots K\}$ is the smallest value which improves the sample loss and satisfies the sample KL-divergence constraint.
        Fit value function by regression on mean-squared error:
        $$
        \phi_{k+1}=\arg \min _{\phi} \frac{1}{\left|\mathcal{D}_{k}\right| T} \sum_{\tau \in \mathcal{D}_{k}} \sum_{t=0}^{T}\left(V_{\phi}\left(s_{t}\right)-\hat{R}_{t}\right)^{2}
        $$
        typically via some gradient descent algorithm.
        
    }
\end{algorithm}

\subsection{Proximal Policy Optimization (PPO)}
As stated in the original paper \cite{article}, PPO is an algorithm that achieves the data efficiency and reliability of TRPO while utilizing just first-order optimization. PPO provides a unique objective function with clipped probability ratios that gives a pessimistic assessment (i.e., lower limit) of the policy's performance. PPO alternates between collecting data from the policy and executing many epochs of optimization on the sampled data \cite{article} to optimize policies.

PPO is a novel policy gradient technique family that alternates between sampling data through interaction with the environment and maximizing a "surrogate" objective function using stochastic gradient ascent. Unlike traditional policy gradient approaches, which conduct one gradient update per data sample, PPO employs an updated objective function that allows for several epochs of mini-batch updates \cite{article}.

This objective function of PPO can be represented as \cite{article}:
\begin{equation}
    \mathcal{L}(s,a,\theta_k, \theta) =  \min\left( \frac{\pi_\theta(a \vert s)}{\pi_{\theta_{old}}(a \vert s)} A^{\pi\theta_{k}}(s, a),
    g(\epsilon, A^{\pi\theta_{k}}(s, a)\right)
    \label{PPO Objective Function 1} 
\end{equation}

where

\begin{equation}
    g(\epsilon, A)= \left\{ 
          \begin{array}{ c l }
            (1 + \epsilon)A & \quad \textrm{if } A \geq 0 \\
            (1 - \epsilon)A & \quad \textrm{if } {A < 0}
          \end{array}
        \right.
\end{equation}

In the implementation, PPO maintains two policy networks. The first one is the current policy that needs to be refined \cite{article}:

\begin{equation}
    \pi_{\theta}(a_t \vert s_t)
\end{equation}

The policy that was used last to collect samples:

\begin{equation}
    \pi_{\theta_k}(a_t \vert s_t)
\end{equation}

PPO switches between sampling data and interacting with the environment. In order to increase sample efficiency, a new policy is reviewed using samples obtained from an earlier policy using the concept of significance sampling. \cite{article}.

\begin{equation}
    \underset{\theta}{\textbf{maximize}} \quad \hat{\mathbb{E}_t} \left[\frac{\pi_\theta(a_t \vert s_t)}{\pi_{\theta_{old}}(a_t \vert s_t)} \hat{A_t}\right]
\end{equation}

As the current policy gets developed, the gap between it and the previous policy grows bigger. The estimation's variance grows, which leads to poor judgments due to inaccuracy. So, say, every four iterations, we resynchronize the second network with the revised policy. \cite{article}.

With clipped objective, we compute a ratio between the new policy and the old policy \cite{article}:

\begin{equation}
        r_{t}(\theta) = \frac{\pi_\theta(a_t \vert s_t)}{\pi_{\theta_{k}}(a_t \vert s_t)} 
        \label{ratio between old and new policies PPO} 
\end{equation}

This ratio compares the two policies. If the new policy is distant from the previous policy, a new objective function is created to clip the estimated advantage function. The new objective function is now \cite{article}:

\begin{equation}
    \mathcal{L}^{CLIP}_{\theta_k} (\theta) = \underset{\tau \sim \pi_k}{\mathbb{E}} \left[ \sum_{t=0}^{T} \left[ \min( r_t(\theta) \hat{A}_{t}^{\pi_{k}}, {clip}(r_t(\theta), 1 - \epsilon, 1 + \epsilon) \hat{A}_{t}^{\pi_{k}}) \right] \right]
    \label{PPO objective function 3}
\end{equation}

The advantage function will be trimmed if the probability ratio between the new and old policies goes beyond the range $(1 - \epsilon)$ and $(1 + \epsilon)$. The clipping restricts the amount of effective change one may make at each phase to increase stability. This inhibits major policy changes if they are outside of our comfort zone \cite{article}. As a result, the method can be expressed as shown in algorithm \ref{alg:ppo}

\begin{algorithm}[H]
    \caption{PPO with Clipped Objective \protect \cite{article}}\label{alg:ppo}
    \SetKwInOut{Input}{input}
    \Input{initial policy parameters $\theta_{0}$, clipping threshold $\epsilon$}
    \For{$k = 0, 1, 2, ...$}{
        Collect set of partial trajectories $\mathcal{D}_k$ on policy $\pi_k = \pi(\theta_k)$
        
        Estimate advantages $\hat{A}_t^{\pi_k}$ using any advantage estimation algorithm
        
        Compute policy update
        
        \hspace{25pt} $\theta_{k+1} = \arg \max\limits_\theta \mathcal{L}_{\theta_k}^{CLIP}(\theta)$
        
        by taking $K$ steps of minibatch SGD (via Adam), where
        
        \hspace{25pt} $\mathcal{L}^{CLIP}_{\theta_k} (\theta) = \underset{\tau \sim \pi_k}{\mathbb{E}} \left[ \sum\limits_{t=0}^{T} \left[ \min( r_t(\theta) \hat{A}_{t}^{\pi_{k}}, {clip}(r_t(\theta), 1 - \epsilon, 1 + \epsilon) \hat{A}_{t}^{\pi_{k}}) \right] \right]$
        
    }
\end{algorithm}

\chapter{Experiments}
\section{Data - Dow Jones 30}
The Dow Jones 30, also known as Dow Jones Industrial Average (DJIA), refers to the thirty blue-chip publicly-traded U.S companies\footnote{Dow Jones stock companies are 3M, American Express, Amgen, Apple, Boeing, Caterpillar, Chevron, Cisco Systems, Coca-Cola, Disney, Dow, Goldman Sachs, Home Depot, Honeywell, IBM, Intel, Johnson and Johnson, JP Morgan Chase, McDonald’s, Merck, Microsoft, Nike, Procter \& Gamble, Salesforce, Travelers, UnitedHealth, Visa, Walgreens, and Walmart.}. Daily stock data from January 2011 up to  November 2021 (10 year period) was extracted and used. 70 percent of the data was used for training while 30 percent was used for testing the RL agents.

\section{Experiment Scope}
To conduct this study, we carried out experiments involving the eight agents described in chapter 6 using the environment described in chapter 4. We carried out three training runs and a hundred test runs. We then stored the peak and mean performances of each agent for analysis. We also ran the training for 10,000 and 100,000 timesteps for all the RL agents. The rest of this section describes what parameters \& hyperparameters were used for both the environment and the agents.

\subsection{Environment Parameters}
\begin{enumerate}
    \item \textbf{Reward Function} \newline
    In the experiments, we used both reward functions available in the environment - log returns and differential Sharpe ratio.
    
    \item \textbf{Lookback Period} \newline
    The lookback period is the duration the agent observes the environment before taking action. A lookback period of 64 days was used. This was determined based on literature and recommendations from the industry advisors.
    \vspace*{1em}
        
    \item \textbf{Trading Costs} \newline
    We experimented with three different trading costs scenarios - no trading costs, 0.1\% of the stock's price, and 1\% of the stock's price.
    \vspace*{1em}
\end{enumerate}

\subsection{RL Agent Hyper-parameters}
We enumerate the parameters for each of the RL agents below.
\begin{enumerate}
    \item{Normalized Advantage Function (NAF)}
    \begin{itemize}
        \item {layer size: 256}
        \item {batch size: 128}
        \item {buffer size: 10,000}
        \item {LR: 1e-3}
        \item {TAU: 1e-3}
        \item {GAMMA (discount factor): 0.99}
        \item {update\_every: 2}
        \item {number\_of\_updates: 1}
        \item {Seed: 0}
    \end{itemize}
    \vspace*{1em}
    
    \item{REINFORCE}
    \begin{itemize}
        \item {Discount Factor (gamma): 0.99}
        \item {hidden size for linear layers: 128}
    \end{itemize}
    \vspace*{1em}

    \item{Deep Deterministic Policy Gradient (DDPG)}
    \begin{itemize}
        \item {memory capacity: 10000}
        \item {num\_memory\_fill\_episodes: 10}
        \item {gamma (discount factor): 0.99}
        \item {tau: 0.005}
        \item {sigma: 0.2}
        \item {theta: 0.15}
        \item {actor\_lr: 1e-4}
        \item {critic\_lr: 1e-3}
        \item {batch\_size: 64}
        \item {warmup\_steps: 100}
    \end{itemize}
    \vspace*{1em}

    \item{Twin Delayed Deep Deterministic Policy Gradient (TD3)}
    \begin{itemize}
        \item {hidden\_dim: 256}
        \item {memory\_dim: 100,000}
        \item {max\_action: 1}
        \item {discount: 0.99}
        \item {update\_freq: 2}
        \item {tau: 0.005}
        \item {policy\_noise\_std: 0.2}
        \item {policy\_noise\_clip: 0.5}
        \item {actor\_lr: 1e-3}
        \item {critic\_lr: 1e-3}
        \item {batch\_size: 128}
        \item {exploration\_noise: 0.1}
        \item {num\_layers: 3}
        \item {dropout: 0.2}
        \item {add\_lstm: False}
        \item {warmup\_steps: 100}
    \end{itemize}
    \vspace*{1em}
    
    \item{Advantage Actor Critic (A2C)}
    \begin{itemize}
        \item {hidden\_dim: 256}
        \item {entropy\_beta: 0}
        \item {gamma (discount factor): 0.9}
        \item {actor\_lr: 4e-4}
        \item {critic\_lr: 4e-3}
        \item {max\_grad\_norm: 0.5}
    \end{itemize}
    \vspace*{1em}
    
    \item{Soft Actor Critic(SAC)}
    \begin{itemize}
        \item {hidden\_dim: 256}
        \item {value\_lr: 3e-4}
        \item {soft\_q\_lr: 3e-4}
        \item {policy\_lr: 3e-4}
        \item {gamma (discount factor): 0.99}
        \item {mean\_lambda: 1e-3}
        \item {std\_lambda: 1e-3}
        \item {z\_lambda: 0.0}
        \item {soft\_tau: 1e-2}
        \item {replay\_buffer\_size: 1,000,000}
        \item {batch\_size: 128}
    \end{itemize}
    \vspace*{1em}
    
    \item{Trust Region Policy Optimization (TRPO)}
    \begin{itemize}
        \item {damping: 0.1}
        \item {episode\_length: 2000}
        \item {fisher\_ratio: 1}
        \item {gamma (discount factor): 0.995}
        \item {l2\_reg: 0.001}
        \item {lambda\_: 0.97}
        \item {lr (learning-rate): 0.001}
        \item {max\_iteration\_number: 200}
        \item {max\_kl (kl-divergence): 0.01}
        \item {val\_opt\_iter: 200}
        \item {value\_memory:1}
    \end{itemize}
    \vspace*{1em}
    
    \item{Proximal Policy Optimization (PPO)}
    \begin{itemize}
        \item {timesteps\_per\_batch: 50,000}
        \item {max\_timesteps\_per\_episode: 2,000}
        \item {n\_updates\_per\_iteration: 5}
        \item {lr (learning-rate): 0.005}
        \item {gamma (discount factor): 0.95}
        \item {clip: 0.2}
    \end{itemize}  
    \vspace*{1em}
\end{enumerate}

\section{Metrics}
In this work, we use the following metrics when backtesting to evaluate and compare the performance of the RL agents:

\begin{enumerate}
    \item{\textbf{Annualized Returns}}\\
    This is the yearly average profit from the trading strategy.
    $$
    (1+\text { Return })^{\wedge}(1 / N)-1=\text { Annualized Return }
    $$
    
    where:
    
    N = Number of periods measured
        
    \item{\textbf{Cumulative Return}}\\
    This is the sum of returns obtained from a trading strategy over a period given an initial investment is known as the cumulative return.
    $$
    Cumulative Return=\left(P_{\text {current }}-P_{\text {initial }}\right) / P_{\text {initial }}
    $$
    
    where:
    
    $P_{\text {current }}=$ Current Price
    
    $P_{\text {initial }}=$ Original Price
    
    \item{\textbf{Sharpe Ratio}}\\
    This is the reward/risk ratio or risk-adjusted rewards of the trading strategy.
    $$
    Sharpe Ratio =\frac{R_{p}-R_{f}}{\sigma_{p}}
    $$
    Where:
    
    $\mathrm{R}_{\mathrm{p}}=$ return of portfolio
    
    $\mathrm{R}_{\mathrm{f}}=$ risk-free rate
    
    $\sigma_{p}=$ standard deviation of the portfolio's excess return
    
    \item{\textbf{Maximum Drawdown (Max DD)}}\\
    This is the difference between the maximum and minimum values of a portfolio over a time horizon used to measure the downside risk of a trading strategy. It is usually represented as a percentage, and lower values indicate good performance.
    $$
    M D D=\frac{\text { Trough Value - Peak Value }}{\text { Peak Value }}
    $$
    
    \item{\textbf{Calmar Ratio}}\\
    This measures the trading strategy's performance relative to its risk. It is calculated by dividing the average annual rate of return by the maximum drawdown. Similar to the Sharpe ratio, higher values indicate better risk-adjusted performance.
    $$
    \text { Calmar Ratio }=\frac{\mathrm{R}_{\mathrm{p}}-\mathrm{R}_{\mathrm{f}}}{\text { Maximum Drawdown }}
    $$
    Where:
    
    $R_{p}=$ Portfolio return
    
    $\mathrm{R}_{\mathrm{f}}=$ Risk-free rate
    
    $R_{p}-R_{f}=$ Annual rate of return
\end{enumerate}

\chapter{Results \& Discussion}
\section{Results}
\begin{figure}[!htb]
    \centering
    \includegraphics[scale=0.5]{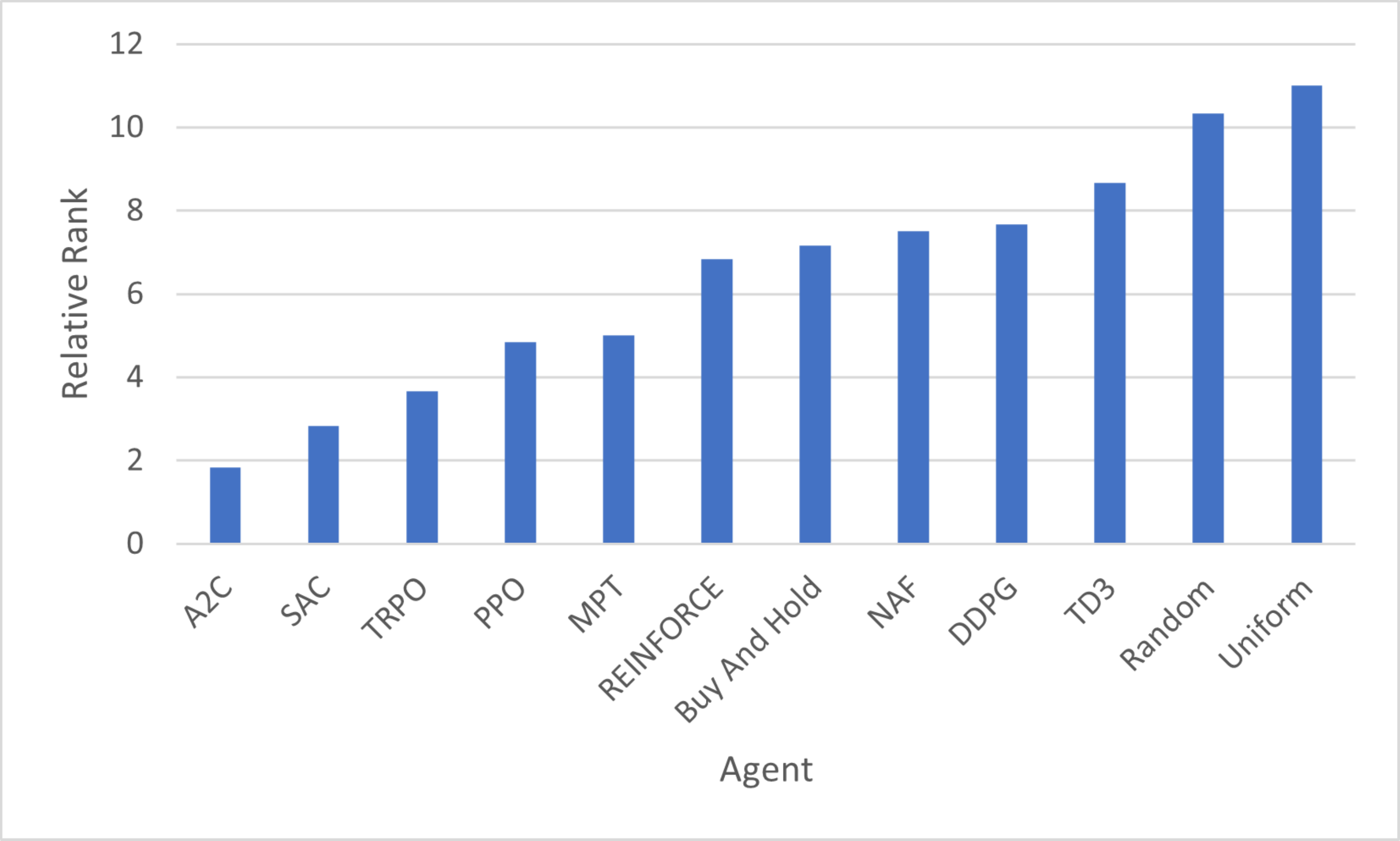}
    \caption{Graph of Final Average Rank of All Agents}
    \label{fig:final-rank}
\end{figure}

\begin{figure}[!htb]
    \centering
    \includegraphics[scale=0.6]{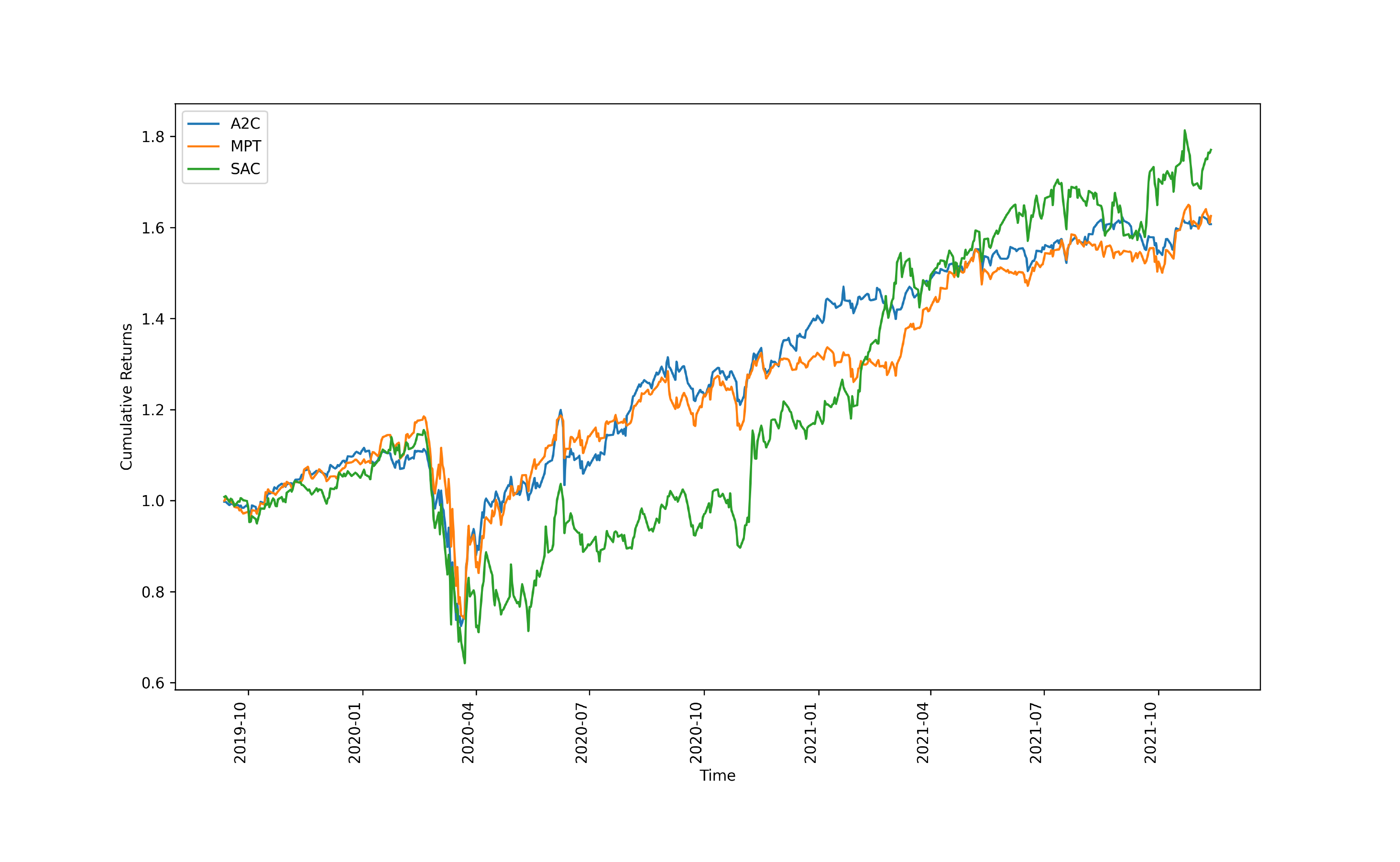}
    \caption{Graph of Cumulative Returns Plot at No Trading Costs }
    \label{fig:CR-0TR-LR}
\end{figure}

\begin{figure}[!htb]
    \centering
    \includegraphics[scale=0.6]{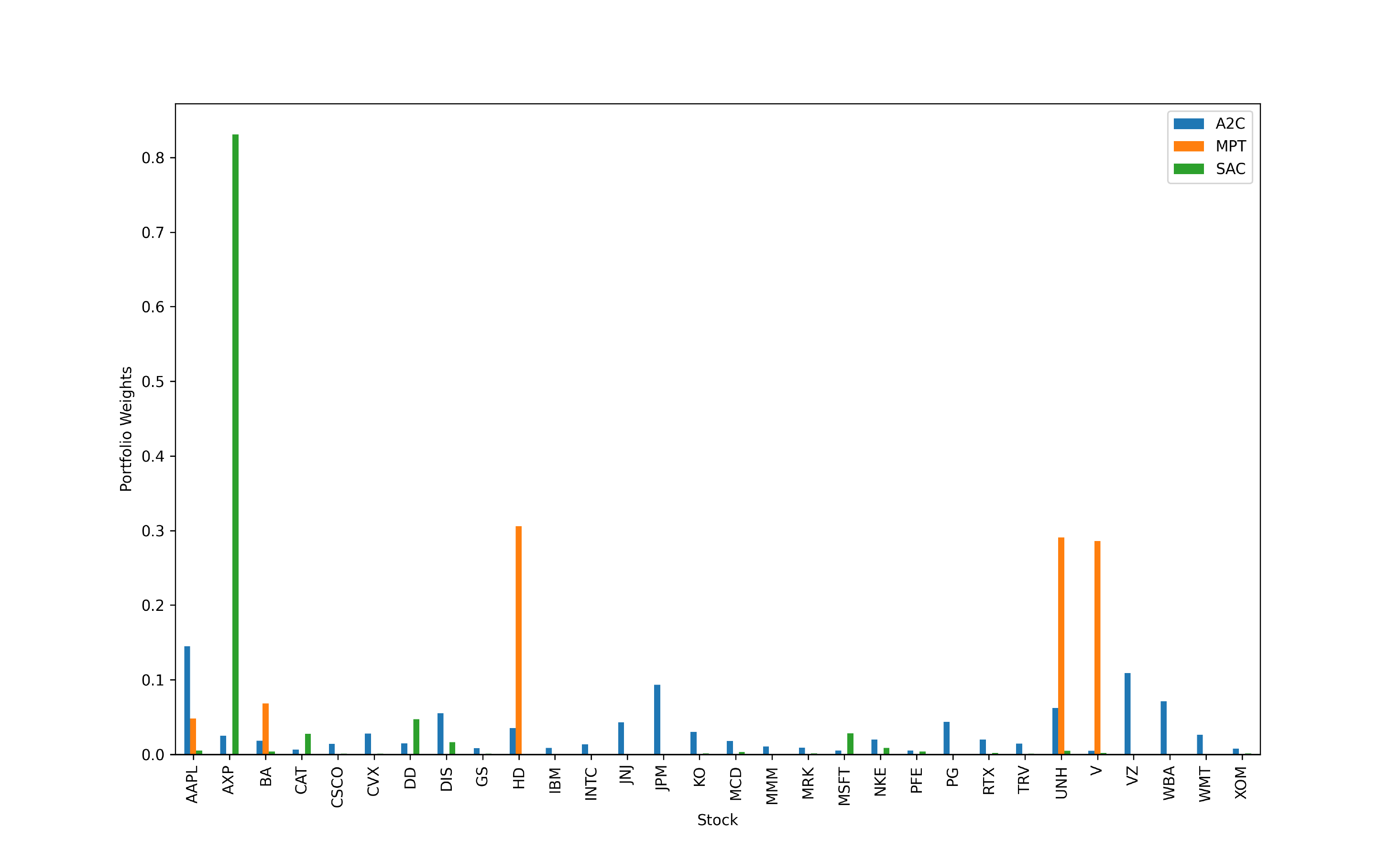}
    \caption{Graph of Mean of Portfolio Weights For Each Stock at No Trading Costs}
    \label{fig:MN-0TR-LR}
\end{figure}

\begin{figure}[!htb]
    \centering
    \includegraphics[scale=0.6]{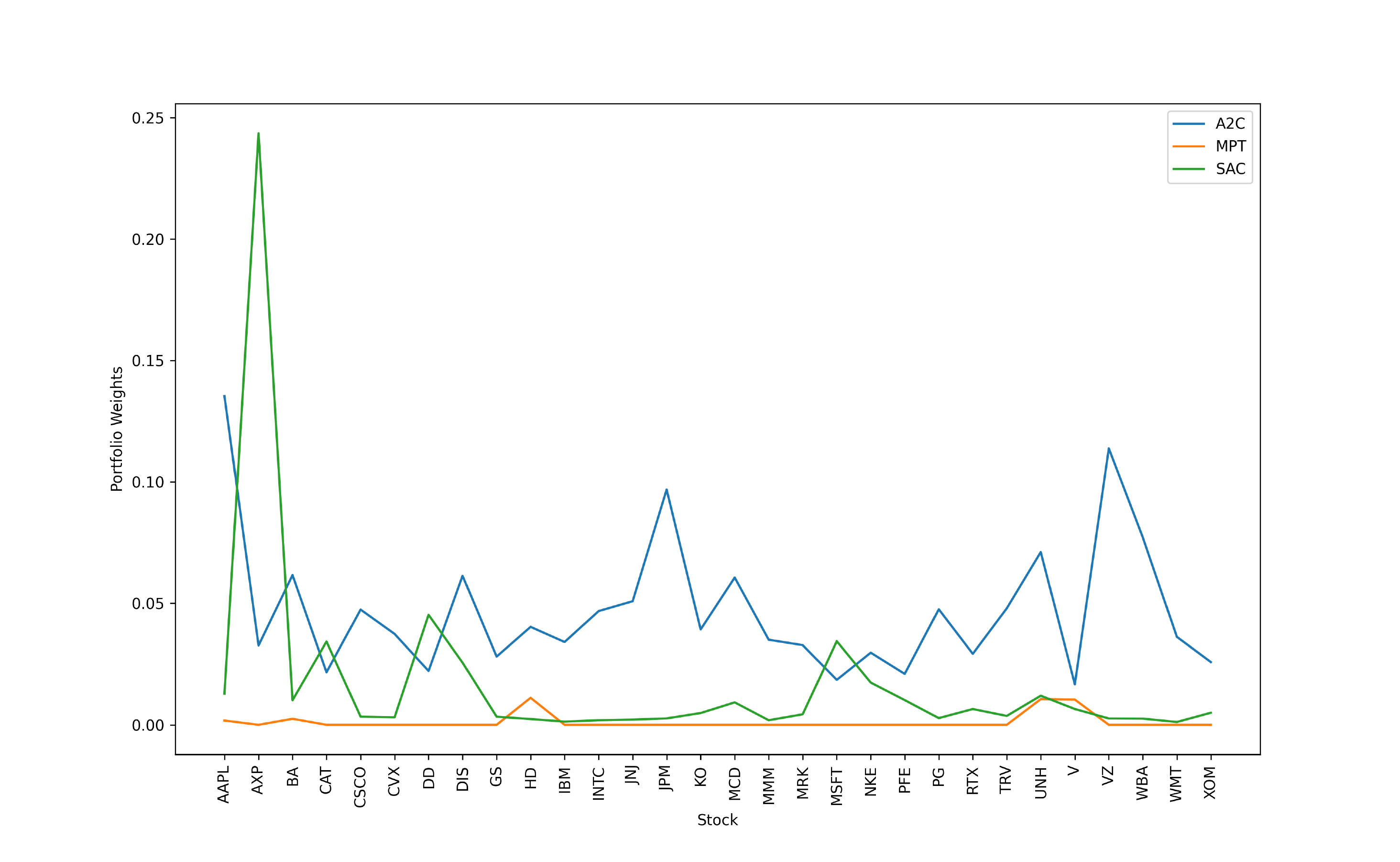}
    \caption{Graph of Mean of Portfolio Weights For Each Stock at No Trading Costs }
    \label{fig:SD-0TR-LR}
\end{figure}

\begin{figure}[!htb]
    \centering
    \includegraphics[scale=0.6]{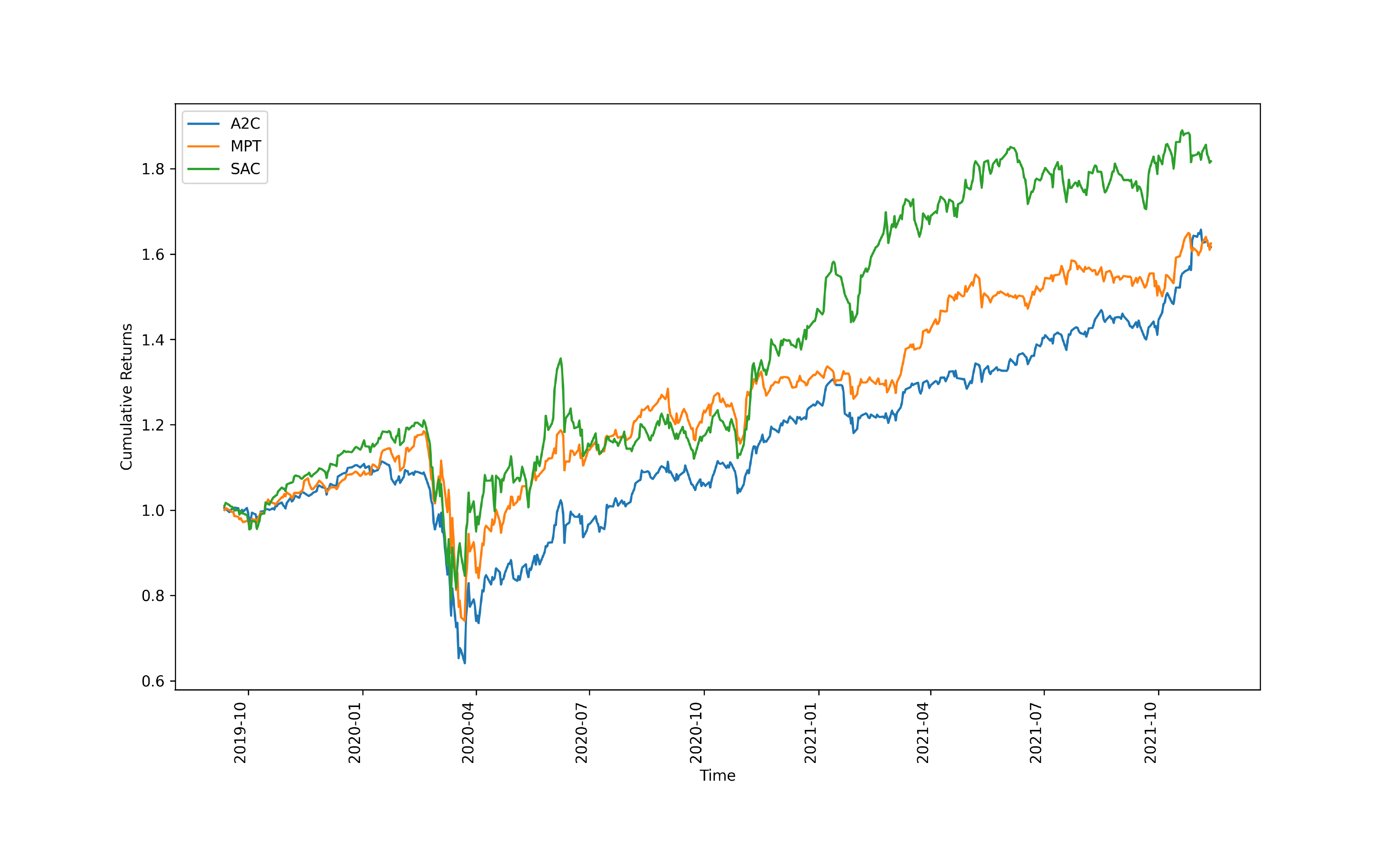}
    \caption{Graph of Cumulative Returns Plot at 0.6\% Trading Costs }
    \label{fig:CR-0.6TR-LR}
\end{figure}

\begin{figure}[!htb]
    \centering
    \includegraphics[scale=0.6]{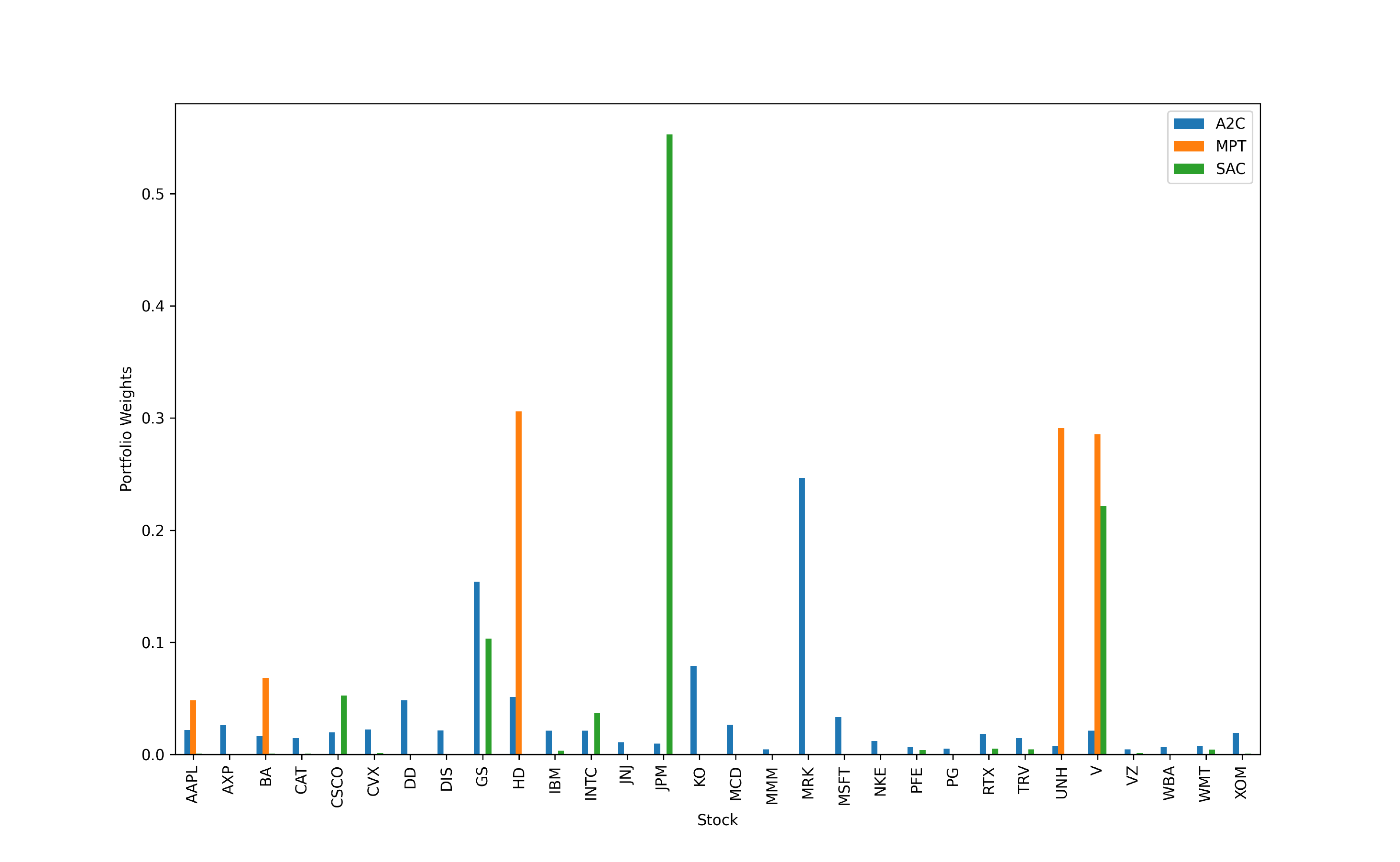}
    \caption{Graph of Mean of Portfolio Weights For Each Stock at 0.6\% Trading Costs }
    \label{fig:MN-0.6TR-LR}
\end{figure}

\begin{figure}[!htb]
    \centering
    \includegraphics[scale=0.6]{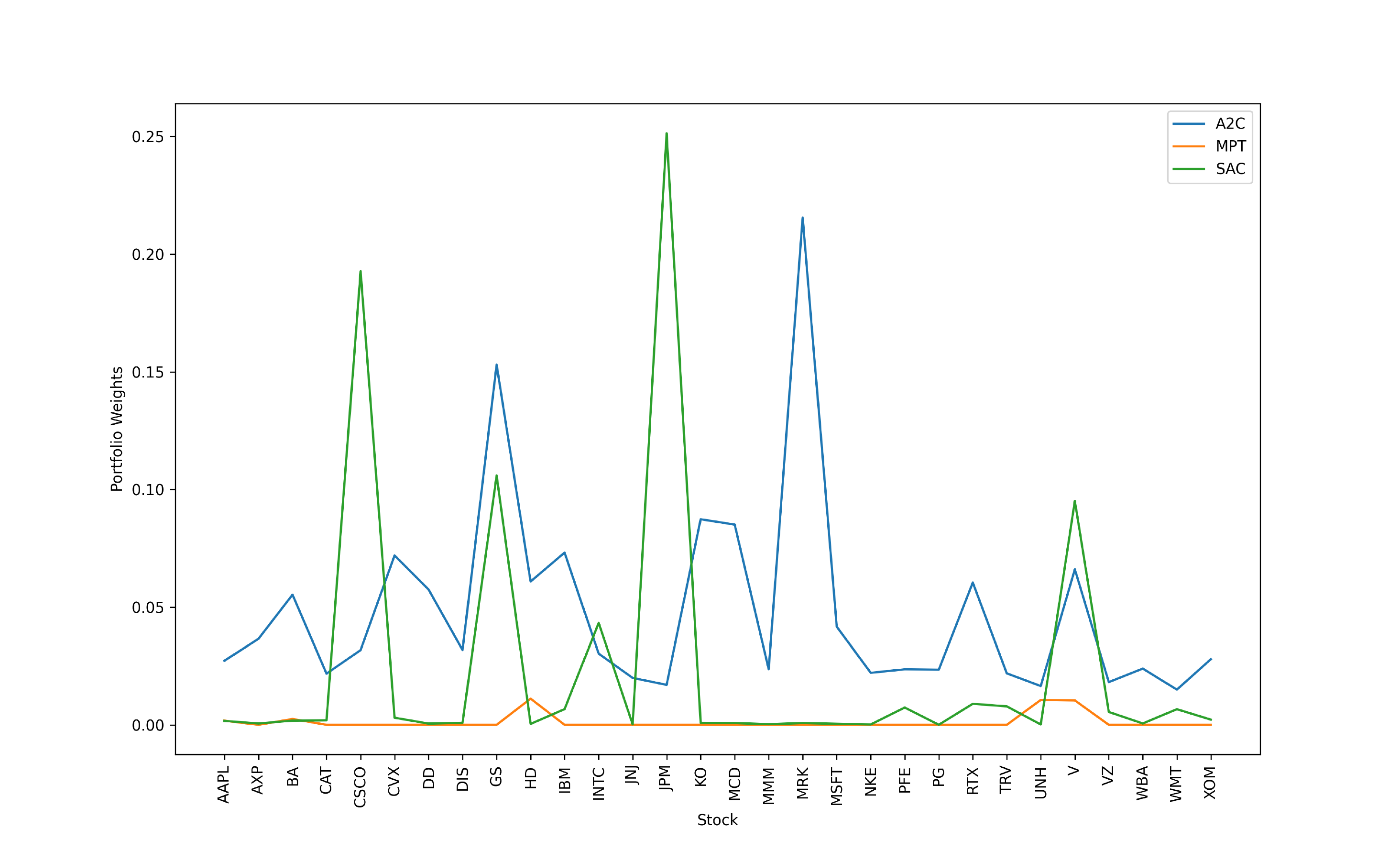}
    \caption{Graph of Mean of Portfolio Weights For Each Stock at 0.6\% Trading Costs }
    \label{fig:SD-0.6TR-LR}
\end{figure}

\begin{figure}[!htb]
    \centering
    \includegraphics[scale=0.6]{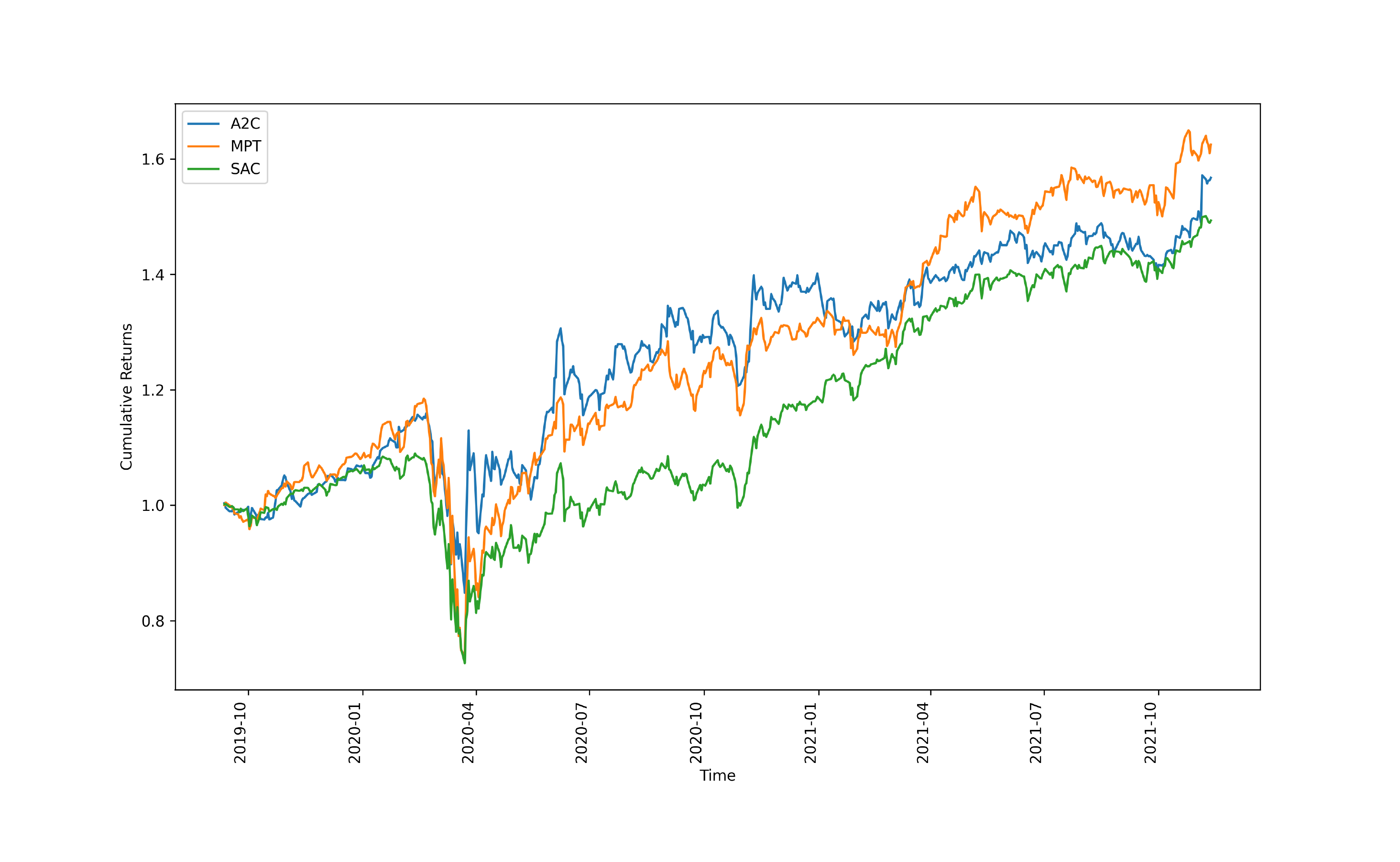}
    \caption{Graph of Cumulative Returns Plot at 1\% Trading Costs }
    \label{fig:CR-1TR-LR}
\end{figure}

\begin{figure}[!htb]
    \centering
    \includegraphics[scale=0.6]{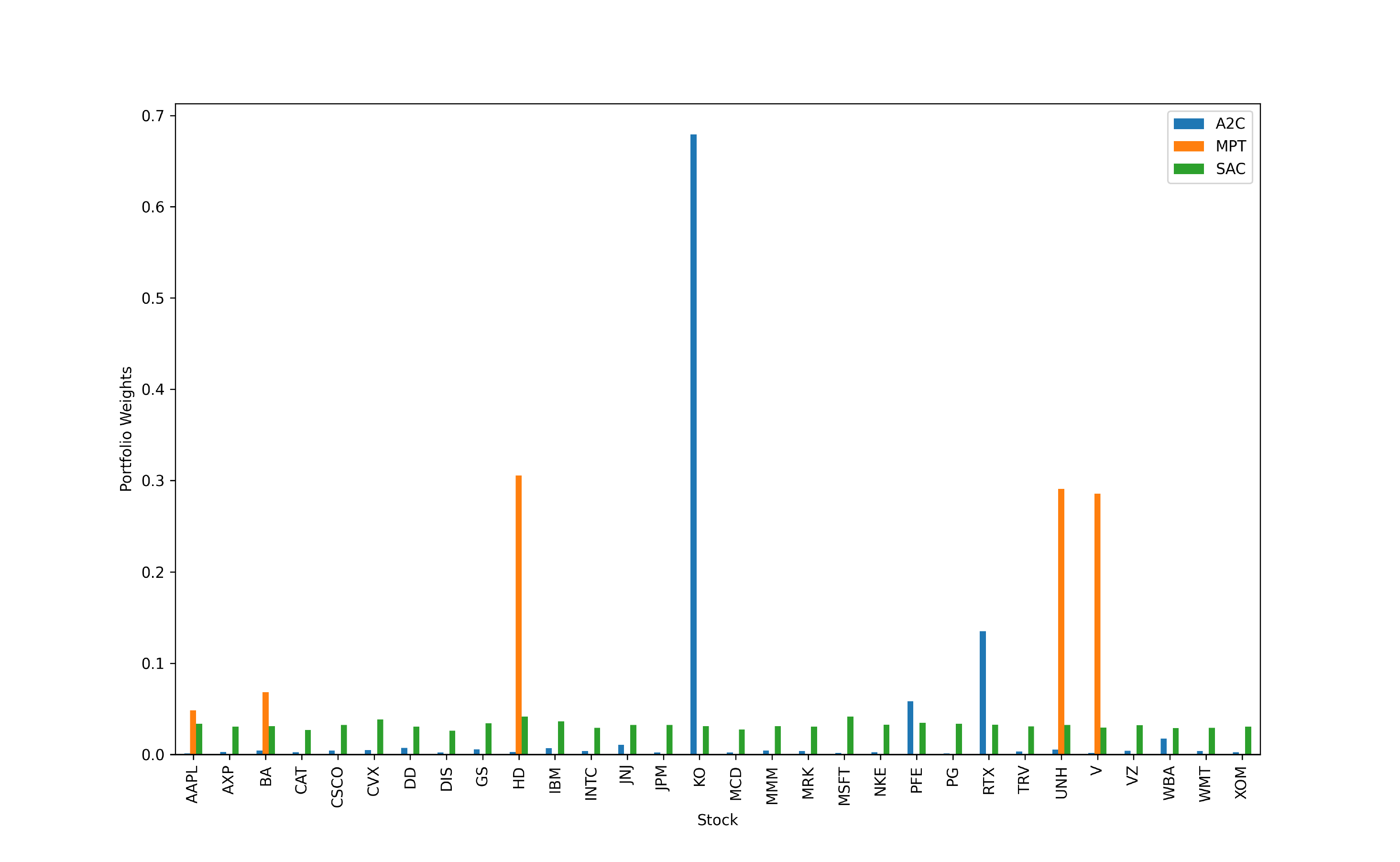}
    \caption{Graph of Mean of Portfolio Weights For Each Stock at 1\% Trading Costs }
    \label{fig:MN-0=1TR-LR}
\end{figure}

\begin{figure}[!htb]
    \centering
    \includegraphics[scale=0.6]{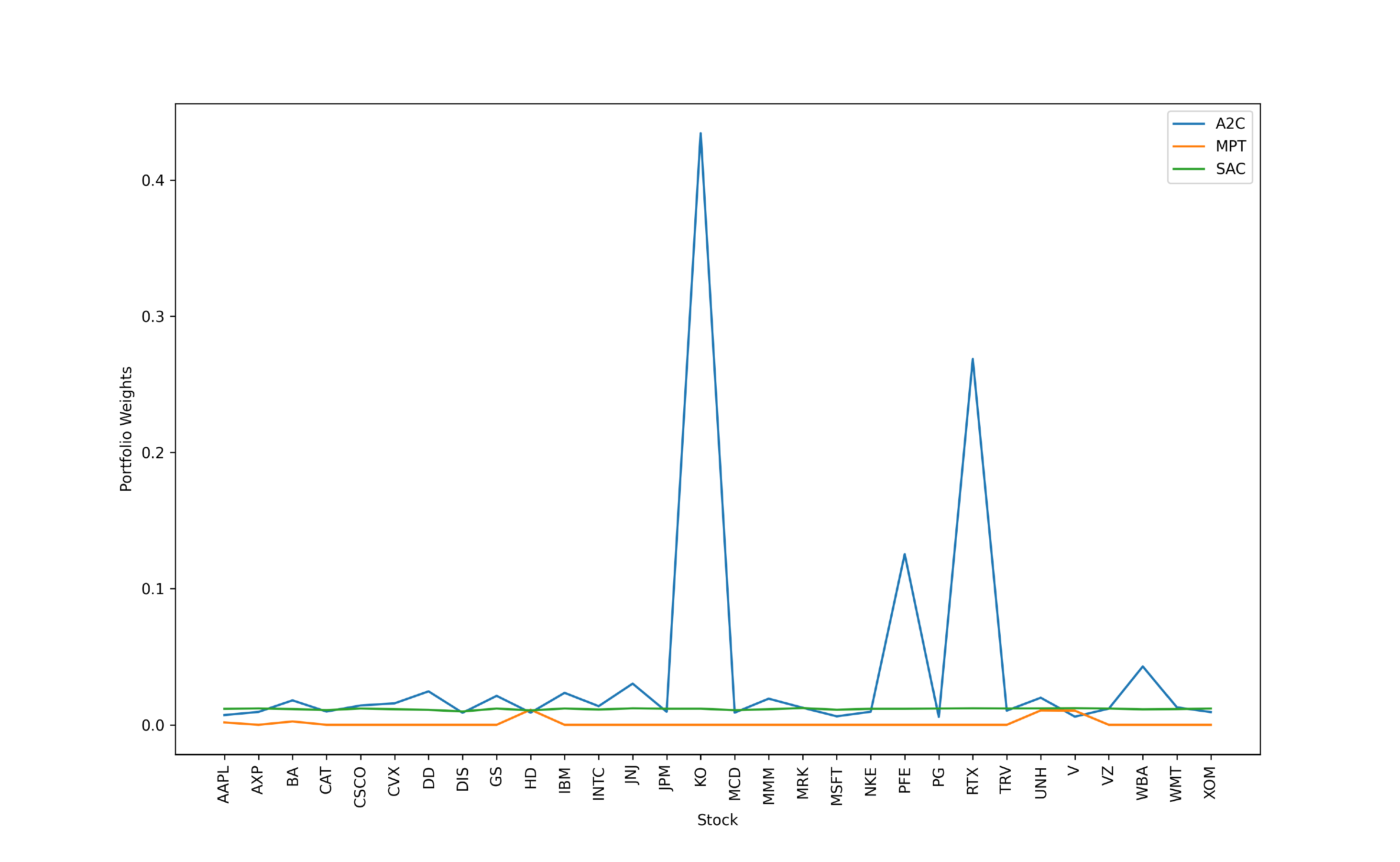}
    \caption{Graph of Mean of Portfolio Weights For Each Stock at 1\% Trading Costs }
    \label{fig:SD-1TR-LR}
\end{figure}

This chapter presents the results from the experiments described in chapter 7. Tables \ref{tab:rank0} to \ref{tab:rank1} present a comparison between the RL agents' mean and peak performance ranks at different trading costs across both reward functions. A huge variation between mean and peak performance shows that an agent's portfolio management strategy is unstable. Figure \ref{fig:final-rank} shows the final average position across all experiments of all the agents from best to worse. Appendix \ref{metrics} shows the raw metric values aggregated into Figure \ref{fig:final-rank}. Figures \ref{fig:CR-0TR-LR} to \ref{fig:SD-1TR-LR} show plots of cumulative returns and portfolio management strategies for the best performing baseline model (MPT) and the RL agents (A2C and SAC) that consistently outperformed MPT based on the average rank metric in  Tables \ref{tab:rank0} to \ref{tab:rank1}. The mean of portfolio weights graphs provide information on how an agent distributes its portfolio among the available stocks while the standard deviation of portfolio weights graphs provide information about how much an agent changes its portfolio distribution. Together, these graphs explain an agent's portfolio management strategy.

\begin{table}[H]
    \centering
    \renewcommand*{\arraystretch}{1.2}
    \caption{Table of Rank Comparison at No Trading Cost}
    \hspace*{-1.5cm}\begin{tabular}{|c|c|c|c|c|}
    \hline
        ~ & Peak Performance Rank & Mean Performance Rank & Difference & Avg Rank \\ \hline
        TRPO & 1 & 2 & 1 & 1.5 \\ \hline
        SAC & 3 & 1 & 2 & 2 \\ \hline
        A2C & 5 & 2 & 3 & 3.5 \\ \hline
        MPT & 6 & 4 & 2 & 5 \\ \hline
        PPO & 4 & 7 & 3 & 5.5 \\ \hline
        REINFORCE & 2 & 11 & 9 & 6.5 \\ \hline
        DDPG & 7 & 8 & 1 & 7.5 \\ \hline
        TD3 & 10 & 5 & 5 & 7.5 \\ \hline
        NAF & 7 & 9 & 2 & 8 \\ \hline
        Buy And Hold & 11 & 6 & 5 & 8.5 \\ \hline
        Random & 9 & 12 & 3 & 10.5 \\ \hline
        Uniform & 12 & 10 & 2 & 11 \\ \hline
    \end{tabular}
    \label{tab:rank0}
\end{table}

\begin{table}[H]
    \centering
    \renewcommand*{\arraystretch}{1.2}
    \caption{Table of Rank Comparison at 0.1\% Trading Cost}
    \hspace*{-1.5cm}\begin{tabular}{|c|c|c|c|c|}
    \hline
        ~ & Peak Performance Rank & Mean Performance Rank & Difference & Avg Rank \\ \hline
        A2C & 1 & 1 & 0 & 1 \\ \hline
        SAC & 4 & 2 & 2 & 3 \\ \hline
        TRPO & 2 & 6 & 4 & 4 \\ \hline
        PPO & 5 & 5 & 0 & 5 \\ \hline
        MPT & 7 & 4 & 3 & 5.5 \\ \hline
        Buy And Hold & 10 & 3 & 7 & 6.5 \\ \hline
        NAF & 6 & 8 & 2 & 7 \\ \hline
        REINFORCE & 3 & 11 & 8 & 7 \\ \hline
        DDPG & 8 & 7 & 1 & 7.5 \\ \hline
        TD3 & 11 & 9 & 2 & 10 \\ \hline
        Random & 9 & 12 & 3 & 10.5 \\ \hline
        Uniform & 12 & 10 & 2 & 11 \\ \hline
    \end{tabular}
    \label{tab:rank0.1}
\end{table}

\begin{table}[H]
    \centering
    \renewcommand*{\arraystretch}{1.2}
    \caption{Table of Rank Comparison at 1\% Trading Cost}
    \hspace*{-1.5cm}\begin{tabular}{|c|c|c|c|c|}
    \hline
        ~ & Peak Performance Rank & Mean Performance Rank & Difference & Avg Rank \\ \hline
        A2C & 1 & 1 & 0 & 1 \\ \hline
        SAC & 5 & 2 & 3 & 3.5 \\ \hline
        PPO & 4 & 4 & 0 & 4 \\ \hline
        MPT & 6 & 3 & 3 & 4.5 \\ \hline
        TRPO & 2 & 9 & 7 & 5.5 \\ \hline
        Buy And Hold & 9 & 4 & 5 & 6.5 \\ \hline
        REINFORCE & 3 & 11 & 8 & 7 \\ \hline
        NAF & 7 & 8 & 1 & 7.5 \\ \hline
        DDPG & 10 & 6 & 4 & 8 \\ \hline
        TD3 & 11 & 6 & 5 & 8.5 \\ \hline
        Random & 8 & 12 & 4 & 10 \\ \hline
        Uniform & 12 & 10 & 2 & 11 \\ \hline
    \end{tabular}
    \label{tab:rank1}
\end{table}

\clearpage

\section{Discussion}
\subsection{RL vs. Baselines}
From Tables \ref{tab:rank0} to \ref{tab:rank1}, we see that the only two baseline agents compare favourably with the RL agents. These baseline agents are Buy \& Hold and MPT, with the latter being the stronger one. Trading costs have significant effects on the performance of the trading agents. Every RL agent outperforms Buy \& Hold at no trading costs. However, when trading costs are introduced, only four RL agents outperform Buy \& Hold.  This is still significant achievement and provides evidence that the RL agents are able to discover good portfolio management strategies.

\subsection{Value-Based RL vs. Policy-Based RL}
In this section, we compare the performance of the NAF agent (a value-based agent) against the REINFORCE agent (a policy-based agent). Tables \ref{tab:rank0} to \ref{tab:rank1} show that both NAF and REINFORCE agents are not exceptional performers, usually under-performing compared to the MPT and the Buy \& Hold baseline agents. On mean performance, the NAF agent outperforms the REINFORCE agent by a small margin regardless of what trading costs is used. However, at peak performance, the REINFORCE agent significantly outperforms the NAF agent. This shows how unstable the policy generated by the REINFORCE agent is. This is further illustrated by the fact that the mean and peak performance rank of the NAF agent across all trading costs vary by as most two positions while those of the REINFORCE agent vary by as much as nine positions. The results obtained from these two agents are consistent with the theoretical understanding of how they work. The REINFORCE agent's policies have a high variance because of sample inefficiency caused by policy gradient estimations from rollout. The NAF agent's policies are much more stable, but its performance is sub-optimal, never outperforming two of the baseline agents.

\subsection{On-Policy vs. Off-Policy}
In this project, there are four on-policy agents (A2C, PPO, REINFORCE, and TRPO) and four off-policy agents (DDPG, NAF, SAC, and TD3). In this section, we will be analyzing the performance of these groups of agents. First, we note that the two agents (A2C and SAC) that consistently outperform the MPT baseline belong to both groups. This provides evidence that both on-policy and off-policy RL agents can perform portfolio management. At the mean performance, SAC slightly outperforms A2C at no trading costs, but A2C slightly outperforms SAC when any form of trading costs was introduced. This is also true at peak performance. Since trading costs are usually involved in the real-world, A2C is better suited to real world portfolio management. 

Comparing the holding strategy of both A2C and SAC from figures \ref{fig:CR-0TR-LR} to \ref{fig:SD-1TR-LR}, we see that the strategy changes with the trading cost. At no trading cost, the SAC agent put about 80\% of its stock into the AXP stock on average. Furthermore, it spreads its portfolio primarily across three other stocks (CAT, DD, MSFT) over the entire testing period. However, the A2C agent took a different strategy. It distributed its portfolio over most of the available stocks, and the spread changes by an average of 5\% across all the stocks over the testing period. It should be noted that A2C had similar cumulative returns as the MPT baseline, which also put 90\% of its holdings into just three stocks - HD, UNH, and V. This confirms a general theory in portfolio management - different market strategies could yield similar results.

When a trading cost of 0.1\% is introduced, both A2C and SAC agents change their strategy. Rather than put 80\% of its holdings into one stock only, the SAC agent put a similar percentage into four different stocks (JPM, V, GS, CSCO) and kept its portfolio spread over them. It is interesting to note that these stocks are entirely different from those it chose at no trading costs, and yet, it outperformed itself on most of the metrics. Similarly, rather than spread its portfolio into most of the available stocks, the A2C agent put about half of its portfolio into three stocks (MRK, GS, KO) this time. Nevertheless, it still kept its portfolio spread across all stocks but usually chose to trade one of CVX, GS, KO, MCD, MRK, RTX, or V.

When the trading cost was 1\%, the strategy landscape changed dramatically. The SAC agent chose a buy and hold strategy and held an almost uniform proportion of stocks across all the available stocks. On the other hand, the A2C agent put most of its stocks (about 90\%) into just three stocks - KO, PFE, and RTX. Also, it kept the portfolio spread across just these three stocks. It is necessary to note that while both SAC and A2C underperformed compared to MPT at 1\% trading costs on returns-related metrics, the A2C agent's strategy enabled it to outperform MPT on risk-related metrics and overall, on average.

The performance of on-policy and off-policy agents seem to be similar at mean performance, but on-policy agents consistently significantly outperform off-policy agents at peak performance. Three out of the four on-policy agents (A2C, PPO, TRPO) are ranked in the top five at different trading costs, consistently outperforming the Buy and Hold baseline. In contrast, only one of the four off-policy agents (SAC) ranks consistently in the top 5. The SAC agent's performance can be attributed its maximum entropy learning framework, which allows it to perform stochastic optimization of policies. 

While the performance of SAC shows that off-policy agents can perform as well as on-policy agents in the task of portfolio management, the evidence of this analysis suggests that on-policy agents are more suited to the task of portfolio management in comparison to off-policy agents. The good performance of on-policy agents is because they are better at evaluating policy and sample efficiency is not a significant problem in portfolio management. While unlikely, the off-policy agents may get better with hyperparameter optimization.

\chapter{Conclusion}
\section{Contributions}
This study investigated the performance of RL when applied to portfolio management using model-free deep reinforcement learning agents. We trained several RL agents on real-world stock prices to learn how to perform asset allocation. We compared the performance of these RL agents against some baseline agents. We also compared the RL agents among themselves to understand which classes of agents performed better.

From our analysis, RL agents can perform the task of portfolio management since they significantly outperformed two of the baseline agents (random allocation and uniform allocation). Four RL agents (A2C, SAC, PPO, and TRPO) outperformed the best baseline, MPT, overall. This shows the abilities of RL agents to uncover more profitable trading strategies. 

Furthermore, there were no significant performance differences between value-based and policy-based RL agents. Actor-critic agents performed better than other types of agents. Also, on-policy agents performed better than off-policy agents because they are better at policy evaluation and sample efficiency is not a significant problem in portfolio management.

In summary, this study shows that RL agents can substantially improve asset allocation since they outperform strong baselines. On-policy, actor-critic RL agents showed the most promise based on our analysis. The next section discusses some directions that future works may want to explore to build on this work.

\section{Future Work}
While this work has tried to do a comparative analysis of more RL agents than what is typically available in the literature, we have not exhausted every possible RL agent. A possible extension to this work is applying the same methodology to other potentially useful RL agents and seeing how they perform compared to the analysis done in this report.

Due to time and compute constraints, we have chosen to stay close to the initially proposed hyperparameters seen in the original papers. Thus, another possible extension to this work will be to carry out extensive hyperparameter optimization for all the eight agents studied in this work to see how the performance of these agents changes.

Furthermore, in this project, we focused only on using feedforward neural networks as the function approximator for all the agents as proposed by the initial authors. However, the financial market is a time-series. Using neural networks such as recurrent neural networks, convolutional neural networks, transformers, among others, that can take into account the temporal nature of the market could yield better results.

Finally, while we have used the Dow Jones market as requested by the client, there is potential for comparative analysis across several other markets. It would be interesting to see if the RL agents perform better when the market is small (e.g., just the top 5 technology companies) or large (e.g. the S\&P 500 market). Similarly, other markets from other locations around the world (e.g., the DAX market of Germany, the HK50 market of Hong Kong, and the JSE market of South Africa) can be studied to see how the insights garnered from the Dow Jones market transfer to these new markets.

\appendix
\chapter{Raw Metric Scores for All Experiments}
\label{metrics}
Tables \ref{tab:mn0LR} to \ref{tab:pk1LR} show all the trading agents' mean and peak performances at different reward functions and trading costs. . The rank columns show an algorithm's position, based on its ranks across all the metrics. 

\begin{table}[H]
    \centering
    \caption{Table of Mean Performance at No Trading Cost \& Log Returns Reward}
    \renewcommand*{\arraystretch}{1.3}
    \begin{tabular}{|l|p{0.13\textwidth}|p{0.13\textwidth}|p{0.1\textwidth}|p{0.1\textwidth}|p{0.08\textwidth}|p{0.08\textwidth}|}
    \hline
        ~ & Cumulative Returns & Annualized Return & Sharpe & Calmar & Max DD & Rank \\ \hline
        A2C & 1.61 & 0.24 & 0.92 & 0.69 & 34\% & 2 \\ \hline
        Buy And Hold & 1.48 & 0.2 & 0.82 & 0.58 & 34\% & 6 \\ \hline
        DDPG & 1.41 & 0.17 & 0.76 & 0.52 & 33\% & 8 \\ \hline
        MPT & 1.62 & 0.25 & 0.9 & 0.66 & 38\% & 4 \\ \hline
        NAF & 1.42 & 0.17 & 0.76 & 0.5 & 34\% & 9 \\ \hline
        PPO & 1.43 & 0.18 & 0.79 & 0.54 & 33\% & 7 \\ \hline
        REINFORCE & 1.37 & 0.15 & 0.69 & 0.46 & 33\% & 11 \\ \hline
        Random & 1.4 & 0.17 & 0.72 & 0.47 & 35\% & 12 \\ \hline
        SAC & 1.77 & 0.3 & 0.83 & 0.68 & 33\% & 1 \\ \hline
        TD3 & 1.44 & 0.18 & 0.81 & 0.55 & 33\% & 5 \\ \hline
        TRPO & 1.59 & 0.24 & 0.94 & 0.69 & 34\% & 2 \\ \hline
        Uniform & 1.4 & 0.17 & 0.73 & 0.48 & 35\% & 10 \\ \hline
    \end{tabular}
    \label{tab:mn0LR}
\end{table}

\begin{table}[H]
    \centering
    \caption{Table of Mean Performance at No Trading Cost \& Sharpe Ratio Reward}
    \renewcommand*{\arraystretch}{1.3}
    \begin{tabular}{|l|p{0.13\textwidth}|p{0.13\textwidth}|p{0.1\textwidth}|p{0.1\textwidth}|p{0.08\textwidth}|p{0.08\textwidth}|}
    \hline
        ~ & Cumulative Returns & Annualized Return & Sharpe & Calmar & Max DD & Rank \\ \hline
        A2C & 1.35 & 0.15 & 0.68 & 0.44 & 34\% & 12 \\ \hline
        Buy And Hold & 1.48 & 0.2 & 0.82 & 0.58 & 34\% & 2 \\ \hline
        DDPG & 1.41 & 0.17 & 0.76 & 0.52 & 33\% & 6 \\ \hline
        MPT & 1.62 & 0.25 & 0.9 & 0.66 & 38\% & 4 \\ \hline
        NAF & 1.42 & 0.17 & 0.76 & 0.48 & 34\% & 7 \\ \hline
        PPO & 1.4 & 0.17 & 0.73 & 0.48 & 35\% & 9 \\ \hline
        REINFORCE & 1.4 & 0.17 & 0.74 & 0.49 & 34\% & 8 \\ \hline
        Random & 1.4 & 0.17 & 0.72 & 0.47 & 35\% & 11 \\ \hline
        SAC & 1.41 & 0.17 & 0.77 & 0.52 & 33\% & 5 \\ \hline
        TD3 & 1.43 & 0.18 & 0.78 & 0.52 & 33\% & 3 \\ \hline
        TRPO & 1.59 & 0.24 & 0.94 & 0.69 & 34\% & 1 \\ \hline
        Uniform & 1.4 & 0.17 & 0.73 & 0.48 & 35\% & 9 \\ \hline
    \end{tabular}
    \label{tab:mn0SR}
\end{table}

\begin{table}[H]
    \centering
     \caption{Table of Peak Performance at No Trading Cost}
    \renewcommand*{\arraystretch}{1.3}
    \begin{tabular}{|l|p{0.13\textwidth}|p{0.13\textwidth}|p{0.1\textwidth}|p{0.1\textwidth}|p{0.08\textwidth}|p{0.08\textwidth}|}
    \hline
        ~ & Cumulative Returns & Annualized Return & Sharpe & Calmar & Max DD & Rank \\ \hline
        A2C & 1.61 & 0.24 & 0.92 & 0.69 & 34\% & 5 \\ \hline
        Buy And Hold & 1.48 & 0.2 & 0.82 & 0.58 & 34\% & 11 \\ \hline
        DDPG & 1.46 & 0.28 & 0.83 & 0.58 & 33\% & 7 \\ \hline
        MPT & 1.62 & 0.25 & 0.9 & 0.66 & 38\% & 6 \\ \hline
        NAF & 1.5 & 0.21 & 0.87 & 0.64 & 32\% & 7 \\ \hline
        PPO & 1.59 & 0.24 & 0.96 & 0.7 & 33\% & 4 \\ \hline
        REINFORCE & 1.64 & 0.25 & 1.03 & 0.8 & 32\% & 2 \\ \hline
        Random & 1.54 & 0.22 & 0.9 & 0.63 & 35\% & 9 \\ \hline
        SAC & 1.77 & 0.3 & 0.83 & 0.68 & 33\% & 3 \\ \hline
        TD3 & 1.44 & 0.26 & 0.81 & 0.55 & 33\% & 10 \\ \hline
        TRPO & 1.93 & 0.35 & 1.34 & 1.3 & 27\% & 1 \\ \hline
        Uniform & 1.4 & 0.17 & 0.73 & 0.48 & 35\% & 12 \\ \hline
    \end{tabular}
    \label{tab:pk0LR}
\end{table}
\begin{table}[H]
    \centering
    \caption{Table of Mean Performance at 0.1\% Trading Costs \& Log Returns Reward}
    \renewcommand*{\arraystretch}{1.3}
    \begin{tabular}{|l|p{0.13\textwidth}|p{0.13\textwidth}|p{0.1\textwidth}|p{0.1\textwidth}|p{0.08\textwidth}|p{0.08\textwidth}|}
    \hline
        ~ & Cumulative Returns & Annualized Return & Sharpe & Calmar & Max DD & Rank \\ \hline
        A2C & 1.62 & 0.25 & 0.88 & 0.68 & 28\% & 1 \\ \hline
        Buy And Hold & 1.48 & 0.2 & 0.82 & 0.58 & 34\% & 3 \\ \hline
        DDPG & 1.42 & 0.18 & 0.78 & 0.52 & 34\% & 7 \\ \hline
        MPT & 1.62 & 0.25 & 0.9 & 0.66 & 38\% & 4 \\ \hline
        NAF & 1.41 & 0.17 & 0.76 & 0.5 & 34\% & 8 \\ \hline
        PPO & 1.44 & 0.18 & 0.78 & 0.54 & 34\% & 5 \\ \hline
        REINFORCE & 1.36 & 0.15 & 0.69 & 0.46 & 33\% & 11 \\ \hline
        Random & 1.37 & 0.16 & 0.69 & 0.45 & 35\% & 12 \\ \hline
        SAC & 1.82 & 0.31 & 0.99 & 0.9 & 35\% & 2 \\ \hline
        TD3 & 1.4 & 0.17 & 0.74 & 0.49 & 34\% & 9 \\ \hline
        TRPO & 1.42 & 0.18 & 0.76 & 0.53 & 33\% & 6 \\ \hline
        Uniform & 1.4 & 0.17 & 0.73 & 0.48 & 35\% & 10 \\ \hline
    \end{tabular}
    \label{tab:mn0.1LR}
\end{table}

\begin{table}[H]
    \centering
    \caption{Table of Mean Performance at 0.1\% Trading Costs \& Sharpe Ratio Reward}
    \renewcommand*{\arraystretch}{1.3}
    \begin{tabular}{|l|p{0.13\textwidth}|p{0.13\textwidth}|p{0.1\textwidth}|p{0.1\textwidth}|p{0.08\textwidth}|p{0.08\textwidth}|}
    \hline
        ~ & Cumulative Returns & Annualized Return & Sharpe & Calmar & Max DD & Rank \\ \hline
        A2C & 1.5 & 0.2 & 0.82 & 0.68 & 28\% & 2 \\ \hline
        Buy And Hold & 1.48 & 0.2 & 0.82 & 0.58 & 34\% & 3 \\ \hline
        DDPG & 1.43 & 0.18 & 0.78 & 0.52 & 34\% & 6 \\ \hline
        MPT & 1.62 & 0.25 & 0.9 & 0.66 & 38\% & 4 \\ \hline
        NAF & 1.41 & 0.17 & 0.76 & 0.48 & 34\% & 8 \\ \hline
        PPO & 1.44 & 0.18 & 0.78 & 0.54 & 34\% & 5 \\ \hline
        REINFORCE & 1.4 & 0.17 & 0.74 & 0.49 & 34\% & 9 \\ \hline
        Random & 1.37 & 0.16 & 0.69 & 0.45 & 35\% & 12 \\ \hline
        SAC & 1.82 & 0.31 & 0.99 & 0.9 & 33\% & 1 \\ \hline
        TD3 & 1.4 & 0.17 & 0.75 & 0.5 & 34\% & 7 \\ \hline
        TRPO & 1.4 & 0.17 & 0.74 & 0.49 & 34\% & 9 \\ \hline
        Uniform & 1.4 & 0.17 & 0.73 & 0.48 & 35\% & 11 \\ \hline
    \end{tabular}
    \label{tab:mn0.1SR}
\end{table}

\begin{table}[H]
    \centering
    \caption{Table of Peak Performance at 0.1\% Trading Costs}
    \renewcommand*{\arraystretch}{1.3}
    \begin{tabular}{|l|p{0.13\textwidth}|p{0.13\textwidth}|p{0.1\textwidth}|p{0.1\textwidth}|p{0.08\textwidth}|p{0.08\textwidth}|}
    \hline
        ~ & Cumulative Returns & Annualized Return & Sharpe & Calmar & Max DD & Rank \\ \hline
        A2C & 1.97 & 0.36 & 1.2 & 0.91 & 29\% & 1 \\ \hline
        Buy And Hold & 1.48 & 0.2 & 0.82 & 0.58 & 34\% & 10 \\ \hline
        DDPG & 1.48 & 0.2 & 0.86 & 0.6 & 33\% & 8 \\ \hline
        MPT & 1.62 & 0.25 & 0.9 & 0.66 & 38\% & 7 \\ \hline
        NAF & 1.5 & 0.21 & 0.87 & 0.63 & 32\% & 6 \\ \hline
        PPO & 1.6 & 0.24 & 0.97 & 0.73 & 33\% & 5 \\ \hline
        REINFORCE & 1.64 & 0.25 & 1.03 & 0.8 & 32\% & 3 \\ \hline
        Random & 1.49 & 0.2 & 0.84 & 0.59 & 34\% & 9 \\ \hline
        SAC & 1.85 & 0.33 & 1.02 & 0.93 & 35\% & 4 \\ \hline
        TD3 & 1.4 & 0.17 & 0.74 & 0.49 & 34\% & 11 \\ \hline
        TRPO & 1.84 & 0.32 & 1.14 & 0.97 & 33\% & 2 \\ \hline
        Uniform & 1.4 & 0.17 & 0.73 & 0.48 & 35\% & 12 \\ \hline
    \end{tabular}
    \label{tab:pk0.1LR}
\end{table}
\begin{table}[H]
    \caption{Table of Mean Performance at 1\% Trading Costs \& Log Returns Reward}
    \renewcommand*{\arraystretch}{1.3}
    \begin{tabular}{|l|p{0.13\textwidth}|p{0.13\textwidth}|p{0.1\textwidth}|p{0.1\textwidth}|p{0.08\textwidth}|p{0.08\textwidth}|}
    \hline
        ~ & Cumulative Returns & Annualized Return & Sharpe & Calmar & Max DD & Rank \\ \hline
        A2C & 1.57 & 0.23 & 0.86 & 0.85 & 27\% & 1 \\ \hline
        Buy And Hold & 1.48 & 0.2 & 0.82 & 0.58 & 34\% & 4 \\ \hline
        DDPG & 1.4 & 0.17 & 0.75 & 0.51 & 33\% & 6 \\ \hline
        MPT & 1.62 & 0.25 & 0.9 & 0.66 & 38\% & 3 \\ \hline
        NAF & 1.41 & 0.17 & 0.75 & 0.5 & 34\% & 8 \\ \hline
        PPO & 1.43 & 0.18 & 0.8 & 0.57 & 31\% & 4 \\ \hline
        REINFORCE & 1.36 & 0.15 & 0.69 & 0.48 & 32\% & 11 \\ \hline
        Random & 1.35 & 0.15 & 0.68 & 0.41 & 36\% & 12 \\ \hline
        SAC & 1.49 & 0.2 & 0.87 & 0.61 & 33\% & 2 \\ \hline
        TD3 & 1.41 & 0.17 & 0.76 & 0.5 & 34\% & 6 \\ \hline
        TRPO & 1.37 & 0.15 & 0.71 & 0.5 & 31\% & 9 \\ \hline
        Uniform & 1.4 & 0.17 & 0.73 & 0.48 & 35\% & 10 \\ \hline
    \end{tabular}
     \label{tab:mn1LR}
\end{table}

\begin{table}[H]
    \centering
    \caption{Table of Mean Performance at 1\% Trading Costs \& Sharpe Ratio Reward}
    \renewcommand*{\arraystretch}{1.3}
    \begin{tabular}{|l|p{0.13\textwidth}|p{0.13\textwidth}|p{0.1\textwidth}|p{0.1\textwidth}|p{0.08\textwidth}|p{0.08\textwidth}|}
    \hline
        ~ & Cumulative Returns & Annualized Return & Sharpe & Calmar & Max DD & Rank \\ \hline
        A2C & 1.58 & 0.23 & 0.97 & 0.85 & 27\% & 1 \\ \hline
        Buy And Hold & 1.48 & 0.2 & 0.82 & 0.58 & 34\% & 2 \\ \hline
        DDPG & 1.4 & 0.17 & 0.75 & 0.51 & 33\% & 5 \\ \hline
        MPT & 1.62 & 0.25 & 0.9 & 0.66 & 38\% & 3 \\ \hline
        NAF & 1.4 & 0.17 & 0.75 & 0.46 & 36\% & 8 \\ \hline
        PPO & 1.39 & 0.16 & 0.72 & 0.48 & 34\% & 10 \\ \hline
        REINFORCE & 1.4 & 0.17 & 0.62 & 0.37 & 36\% & 11 \\ \hline
        Random & 1.35 & 0.15 & 0.68 & 0.41 & 36\% & 12 \\ \hline
        SAC & 1.43 & 0.18 & 0.79 & 0.54 & 33\% & 4 \\ \hline
        TD3 & 1.41 & 0.17 & 0.76 & 0.5 & 34\% & 6 \\ \hline
        TRPO & 1.39 & 0.16 & 0.71 & 0.5 & 31\% & 9 \\ \hline
        Uniform & 1.4 & 0.17 & 0.73 & 0.48 & 35\% & 7 \\ \hline
    \end{tabular}
    \label{tab:mn1SR}
\end{table}

\begin{table}[H]
    \centering
    \caption{Table of Peak Performance at 1\% Trading Costs \& Sharpe Ratio Reward}
    \renewcommand*{\arraystretch}{1.3}
    \begin{tabular}{|l|p{0.13\textwidth}|p{0.13\textwidth}|p{0.1\textwidth}|p{0.1\textwidth}|p{0.08\textwidth}|p{0.08\textwidth}|}
    \hline
        ~ & Cumulative Returns & Annualized Return & Sharpe & Calmar & Max DD & Rank \\ \hline
        A2C & 1.73 & 0.28 & 1.02 & 1.19 & 24\% & 1 \\ \hline
        Buy And Hold & 1.48 & 0.2 & 0.82 & 0.58 & 34\% & 9 \\ \hline
        DDPG & 1.43 & 0.18 & 0.79 & 0.55 & 33\% & 10 \\ \hline
        MPT & 1.62 & 0.25 & 0.9 & 0.66 & 38\% & 6 \\ \hline
        NAF & 1.5 & 0.2 & 0.86 & 0.63 & 32\% & 7 \\ \hline
        PPO & 1.6 & 0.29 & 0.99 & 0.73 & 33\% & 4 \\ \hline
        REINFORCE & 1.64 & 0.25 & 1.03 & 0.82 & 31\% & 3 \\ \hline
        Random & 1.53 & 0.21 & 0.88 & 0.61 & 35\% & 8 \\ \hline
        SAC & 1.49 & 0.29 & 0.86 & 0.61 & 33\% & 5 \\ \hline
        TD3 & 1.41 & 0.17 & 0.76 & 0.5 & 34\% & 11 \\ \hline
        TRPO & 1.62 & 0.25 & 1.06 & 0.82 & 30\% & 2 \\ \hline
        Uniform & 1.4 & 0.17 & 0.73 & 0.48 & 35\% & 12 \\ \hline
    \end{tabular}
    \label{tab:pk1LR}
\end{table}


\bibliographystyle{apacite}
\bibliography{references}
\end{document}